\documentclass[twocolumn]{aa}
\usepackage{graphicx}
\usepackage{txfonts}

\usepackage{amsmath,amstext}
\usepackage[T1]{fontenc}
\usepackage{hyperref}
\usepackage[figure,figure*]{hypcap}
\usepackage{floatrow}
\usepackage{subcaption}
\usepackage{calc}
\usepackage{soul}
\usepackage{cancel}
\usepackage{stfloats}
\usepackage{afterpage}
\usepackage{float}
\floatstyle{plaintop}
\restylefloat{table}
\usepackage[normalem]{ulem}

\begin{document}

\title{Normalization of the extragalactic background light from high-energy $\gamma$-ray observations}

   \author{B. Biasuzzi
          \inst{1}
          \and
          O. Hervet\inst{2}
          \and
          D. A. Williams\inst{2}
          \and
          J. Biteau \inst{1}
          }

   \institute{Institut de Physique Nucl\'eaire, IN2P3/CNRS, Universit\'e Paris-Sud, Univ. Paris/Saclay, 15 rue Georges Cl\'emenceau, 91406 Orsay, Cedex, France\\
         \and
             Santa Cruz Institute for Particle Physics and Department of Physics, University of California at Santa Cruz, Santa Cruz, CA 95064, USA\\
             }

\abstract
{Extragalactic background light (EBL) plays an important role in cosmology since it traces the history of galaxy formation and evolution. Such diffuse radiation from near-UV to far-infrared wavelengths can interact with $\gamma$-rays from distant sources such as active galactic nuclei (AGNs), and is responsible for the high-energy absorption observed in their spectra. However, probing the EBL from $\gamma$-ray spectra of AGNs is not trivial due to internal processes that can mimic its  effect. Such processes are usually taken into account in terms of curvature of the intrinsic spectrum.  Hence,
an improper choice of parametrization for the latter can seriously affect EBL reconstruction. In this paper, we propose a statistical approach that avoids a priori assumptions on the intrinsic spectral curvature and that, for each source, selects
the best-fit model on a solid statistical basis.
By combining the {\it Fermi}-LAT observations of 490 blazars, we determine the $\gamma$-ray-inferred level of EBL for various state-of-the-art EBL models.
We discuss the EBL level obtained from the spectra of both BL~Lacs and flat spectrum radio quasars (FSRQ) in order to investigate the impact of internal absorption in different classes of objects.
We further scrutinize constraints on the EBL evolution from $\gamma$-ray observations by reconstructing the EBL level in four redshift ranges, up to $z\sim2.5$. The approach implemented in this paper, carefully addressing the question of the modeling of the intrinsic emission at the source, can serve as a solid stepping stone for studies of hundreds of high-quality spectra acquired by next-generation $\gamma$-ray instruments.}


\maketitle

\section{INTRODUCTION}
\label{introduction}
The extragalactic background light (EBL) is the brightest background photon field after the cosmic microwave background (CMB) radiation. Its spectrum, which ranges from 0.1 to 1000$\, \mu$m, is formed by the contribution of stars and accreting compact objects during the whole history of the universe. The direct light emitted in the UV and optical bands builds up the so-called 
cosmic optical background (COB), peaking around $\lambda \sim 1$\,$\mu$m, while the light reprocessed by dust close to the emitter and in the interstellar medium contributes to the cosmic infrared background (CIB), peaking around $\lambda \sim 100 \,\mu$m \citep{Dwek2013}. The EBL spectrum is hard to measure directly because of the presence of strong foreground emissions, such as the zodiacal light and the light emitted by the Galactic plane, and therefore direct sky photometry usually provides an overestimation of the EBL and leads to large uncertainties \citep{Hauser2001}. On the other hand, lower limits on the EBL spectrum can be obtained by summing up the light emitted by resolved sources. However, this approach is affected by the limited sensitivity of the surveys used, which can result  in missing objects \citep{Levenson2008, Madau2000, Dole2006}, although convergence is now observed in a number of bands \citep{Driver2016}.

Knowledge of the evolution of the EBL density is fundamental to understanding the history of galaxy formation and evolution, the stellar initial mass function, the metallicity evolution, and all the processes in the universe related to thermal energy release. However, local measurements of the EBL spectrum do not provide any information about its evolution with cosmic epochs, and therefore various approaches have been followed to model the EBL density as a function of redshift, $z$. The so-called empirical models, starting from present galaxy luminosity functions of different populations, aim to reconstruct them back in time by assuming a $z$-dependence inferred by fitting the model prediction to the observed galaxy counts \citep{Stecker2006, Franceschini2008, Dominguez2011, Franceschini2017, Franceschini2018}. On the contrary, in phenomenological models, the EBL density is obtained by modeling the emission processes throughout galaxy formation and evolution in the universe. Such approaches \citep{Dwek1998, Razzaque2009, Finke2010} combine the {\it z}-dependence of the star formation rate and different models of population synthesis to obtain the luminosity density at different redshifts. 
Finally, semi-analytic models \citep[\textit{e.g.},][]{Gilmore2012} predict the evolving EBL out to high redshifts using cosmological simulations and incorporating all the most important physical processes that determine galaxy evolution. The derived predictions of such models are then compared to observational quantities such as galaxy morphology, color, spectral energy distribution (SED), and counts. Deriving phenomenological and semi-analytic models is challenging because such models involve several astrophysical processes, sometimes poorly constrained, but they still remain the most physically motivated models with respect to the empirical ones. 

Another way to carry out studies of the EBL is through observations of high-energy spectra of distant sources, such as active galactic nuclei (AGNs). In particular, blazars, AGNs whose relativistic jets point directly towards the observer, are the best candidates.  They are among the most powerful $\gamma$-ray sources in the sky since their luminosity is amplified by a relativistic boosting of a factor $\sim \delta^4$, where $\delta\sim \mathcal{O}(10)$ is the Doppler factor of the source. 
Blazars are characterized by strongly variable nonthermal emission, whose SED shows two peaks: one located at low energies (from infrared to X-rays) that is thought to arise from synchrotron emission, and one located at high energies (in the MeV-TeV range) that is usually interpreted -- in leptonic scenarios -- as inverse-Compton scattering of relativistic electrons on local photon fields \citep{Band1985, Tavecchio1998, Sikora2008}. On the other hand, in hadronic scenarios \citep{Mannheim1993, Aharonian2000}, a protonic component in the jet would contribute to the high-energy peak, especially via proton synchrotron or via the decay of neutral pions on top of inverse-Compton emission. The characteristic observed high-energy spectrum has a smooth and concave shape ({\it i.e.}, downward sloping). A further cutoff in the spectrum is expected if one of the following processes takes places: intrinsic self-absorption, Klein-Nishina suppression, cut-off in the parent population of accelerated particles, or EBL absorption.

The presence of blazars in a wide redshift range makes them ideal candidates to study the EBL evolution with cosmic epochs. Before reaching the observer, some of the $\gamma$-rays emitted by blazars interact with EBL photons, generating electron--positron pairs \citep{Nikishov1962, Gould1967}. This interaction results in a flux decrease at very-high energies (VHE, $E > 100 \,$GeV), a feature that led several authors to attempt to constrain the EBL from the absorbed region in blazar spectra. Even if this technique is independent from direct measurements, it is limited by the fact that the intrinsic emission processes at play in blazars are still not very well understood. Upper limits on the EBL have been estimated by assuming a maximum hardness of AGN spectra in the $\gamma$-ray band \citep{Stecker1992,Mazin2007,Finke2009}, or by assuming no intrinsic curvature at all \citep{Meyer2012,Sinha2014}.

The {\it Fermi}-LAT and the H.E.S.S. collaborations proposed an approach to measure the EBL, where the EBL absorption is scaled by a normalizing factor, $\alpha$, that indicates the agreement between a given EBL model and the $\gamma$-ray data \citep{Ackermann2012,Hess2013}. By combining observations of several blazars and making minimal assumptions on the intrinsic spectra, they derived the scaling factor for different EBL models.

The pair creation process depends both on the energy of the photons and the redshift of the source. A $\gamma$-ray of energy $E$ can interact with EBL photons up to \mbox{$\lambda_{max} \sim 2.4 \, \mu\text{m} \; (E_0/500 \; \text{GeV})(1+z)^2$,} as imposed by the energy threshold of the process. This means that ground- and space-based $\gamma$-ray observations can explore different regions of the EBL spectrum depending on the $\gamma$-ray energy range and redshift they cover. The {\it Fermi}-LAT Collaboration \citep{Ackermann2012} explored the COB peak of the EBL spectrum by using 150 blazar spectra in the redshift range $0<z<1.6$. The H.E.S.S. Collaboration \citep{Hess2013} used 17 blazar spectra up to $z<0.2$ to probe the EBL spectrum from 0.30 to 17$\, \mu$m. In both approaches, a likelihood maximization is performed over the scaling factor in order to compare the best-fit case with the null hypothesis, $\alpha = 0$ ({\it i.e.}, absence of EBL absorption), which was rejected at the $6\sigma$ level in the {\it Fermi} work, and $9 \sigma$ in the H.E.S.S. one.

By following a similar approach to \cite{Ackermann2012} and \cite{Hess2013}, \cite{Ahnen2015} and \cite{Abeysekara2015} placed constraints on the EBL in a different energy range using observations of PKS 1441+25, the second-most distant VHE flat spectrum radio quasar (FSRQ) located at $z=0.939$ during a high-activity state.
In \cite{Ahnen2016}, both MAGIC and {\it Fermi}-LAT data of the gravitationally lensed blazar B2 018+357 (the most distant VHE one at $z=0.954$) were combined to obtain EBL constraints from $0.3$ to $1.1 \,  \mu$m. In \cite{Ahnen2016b}, a flaring state of the BL~Lac 1ES 1011+496 allowed the MAGIC Collaboration to explore the EBL density between $0.24 \, \mu$m and $4.25 \, \mu$m. Finally, \cite{Mazin2017} extended the method used in \cite{Ahnen2016b} to 30 independent energy spectra, derived from eight BL~Lacs and four FSRQs, to derive the scaling factor of the EBL in the redshift range $0.031 < z < 0.944
$ for the EBL model of \cite{Dominguez2011}.

\cite{Armstrong2017} analyzed the {\it Fermi}-LAT spectra of 16 high-redshift sources ($0.847 \le z \le 1.596$) by following the approach of \cite{Ackermann2012}. After extrapolating the intrinsic spectra from their unabsorbed region (where the absorption is less than 0.1\%), they fit the whole spectrum to derive the EBL level.

The most extensive VHE $\gamma$-ray study on the EBL was carried out by \cite{Biteau2015}. They used 86 spectra (from 29 BL Lac objects with reliable redshifts up to $z=0.287$) from ground-based observatories (MAGIC, H.E.S.S., VERITAS, Whipple, ARGO-YBJ, HEGRA, TACTIC, Tibet, and CAT) and from space-based observatories ({\it Fermi}-LAT), together with the EBL local constraints reported in \cite{Dwek2013}. They investigated the $0.26 - 105 \, \mu$m EBL spectrum region, improving also on the method of the {\it Fermi}-LAT and H.E.S.S. Collaborations by introducing a model-independent approach. The hypothesis of absence of absorption was rejected at the $11 \sigma$ level.

The H.E.S.S. Collaboration derived the EBL intensity, independent of a given EBL model, using H.E.S.S. data alone and an approach similar to \cite{Biteau2015}.
By using 21 spectra from eight high-frequency-peaked BL~Lac objects, they determined the EBL level in four wavebands, from $0.25$ to $98.6\,\mu$m \citep{Hess2017}. The EBL measurement that they obtained is preferred to the null hypothesis at the $9.5\sigma$ level.

The MAGIC Collaboration \citep{Acciari2019} used MAGIC and {\it Fermi}-LAT data, with 32 $\gamma$-ray spectra coming from 12 sources, to reconstruct the EBL normalization in the redshift range $0.03 \le z \le 0.944$. 
They also constrained the EBL spectrum in the range 0.18\,-\,100$\, \mu$m, reaching a total uncertainty of 20\% in the range 0.62\,-\,2.24$\, \mu$m.

Finally, \cite{Abdollahi2018} applied the methodology of \cite{Ackermann2012} to a larger sample of 739 AGNs (419 FSRQs and 320 BL Lacs) from the third catalog of AGNs \citep[3LAC,][]{Ackermann2015} and the GRB 080916C located at $z=4.35$. \cite{Abdollahi2018}   reconstructed the EBL intensity from $\sim$0.1 to $\sim$4 \,$\mu$m, and the cosmic $\gamma$-ray horizon up to a redshift $z=3$. Moreover, they used a physical EBL model to infer the optical depth directly from the star formation rate (SFR). The latter was then optimized to reproduce the {\it Fermi}-LAT optical depth data, thus constraining the SFR over 90\% of the cosmic time.

Another way to probe the EBL, especially at high redshift, consists in using only $\gamma$-ray bursts (GRBs) detected by {\it Fermi}-LAT, as done in \cite{Desai2017}. These latter authors combined 22 GRB observations with redshift in the range $0.15 \le z \le 4.35$, disfavoring the hypothesis of no EBL absorption at the $\sim$2.8$\sigma$ level. They obtained constraints on the EBL optical depth for an effective redshift of 1.8.

So far, the results obtained by analyzing the $\gamma$-ray spectra of AGNs and GRBs are in good agreement with the local constraints on the EBL.

By following the approach in \cite{Ackermann2012} and \cite{Abdollahi2018}, while retaining flexibility as to the shape of each intrinsic $\gamma$-ray spectrum as in \cite{Biteau2015}, we determine the scaling factor for different EBL models by analyzing the {\it Fermi}-LAT spectra of many blazars.
After briefly discussing the impact of the choice of the intrinsic-spectrum model in Sect.~\ref{intr_spec_choice}, the source-selection criteria and the full dataset are presented in Sect.~\ref{data_sample}. The analysis method is described in Sect.~\ref{analysis}. Finally, in Sects.~\ref{results} and \ref{discussion}, the results are presented and compared to previous EBL measurements, together with the estimate of systematic uncertainties.

\section{Impact of intrinsic-spectrum assumption }
\label{intr_spec_choice}

\begin{figure}
\subcaptionbox{\label{a}}{\includegraphics[width=0.45\columnwidth,clip=true]{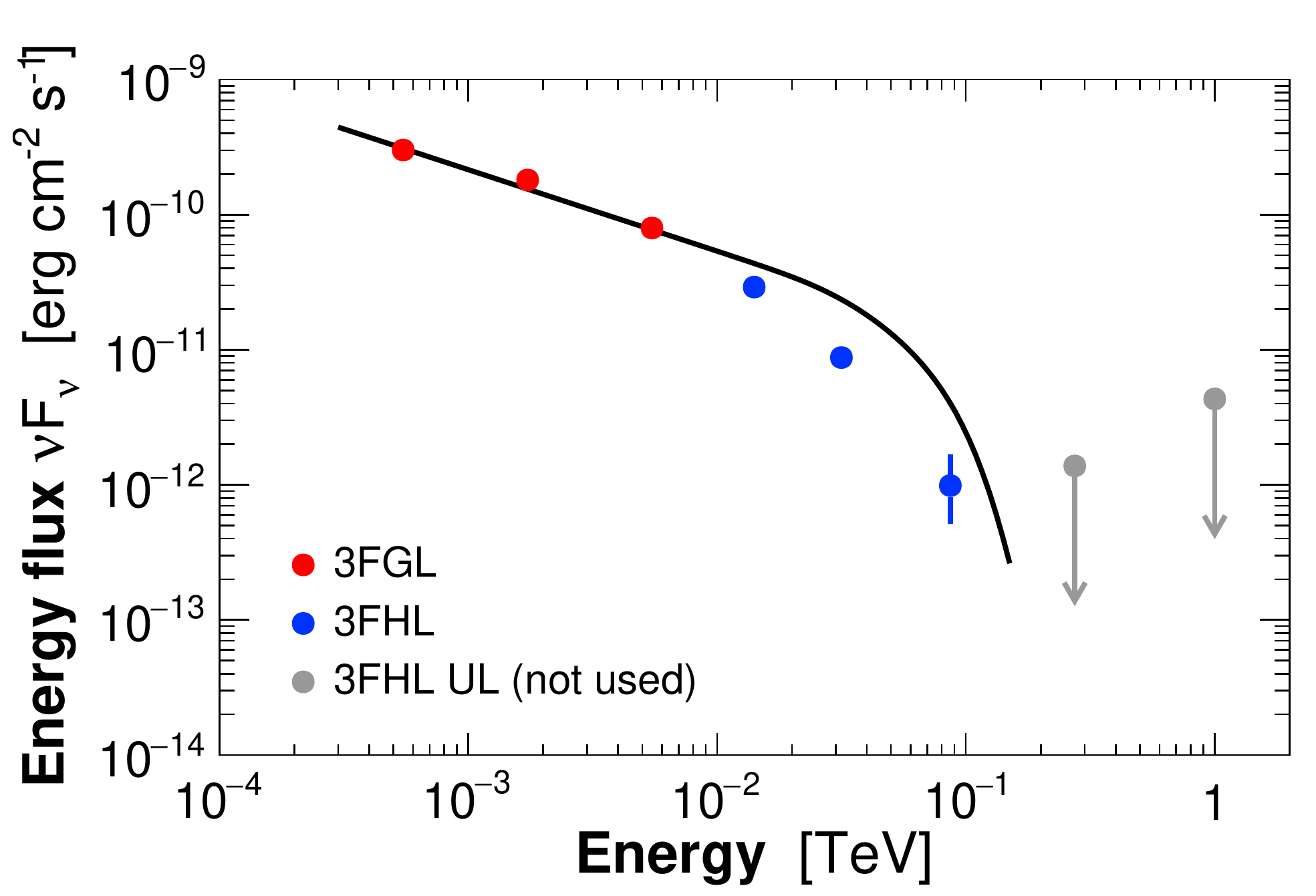}\includegraphics[width=0.45\columnwidth]{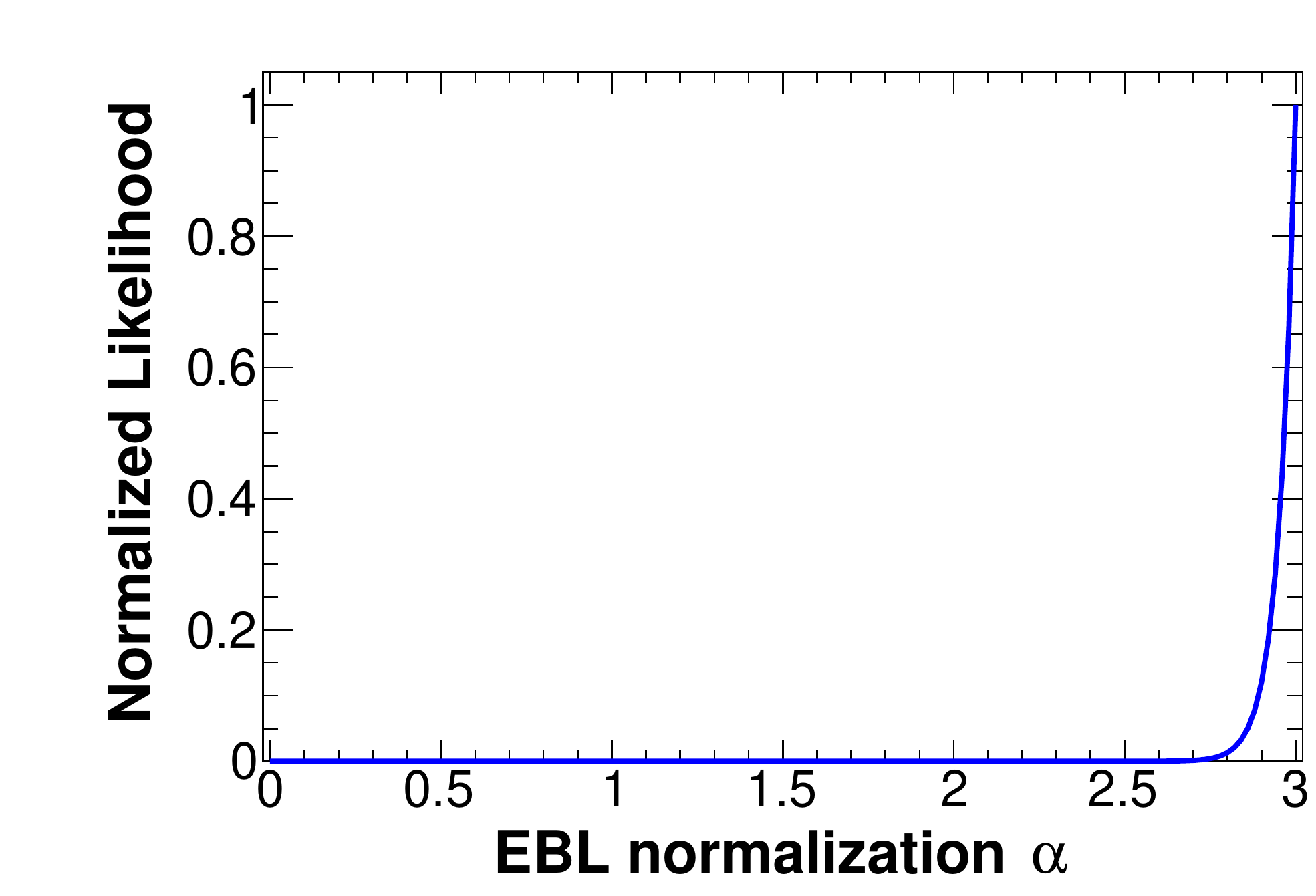}}
\hfill
\hspace{-3mm}
\subcaptionbox{\label{c}}{\includegraphics[width=0.45\columnwidth]{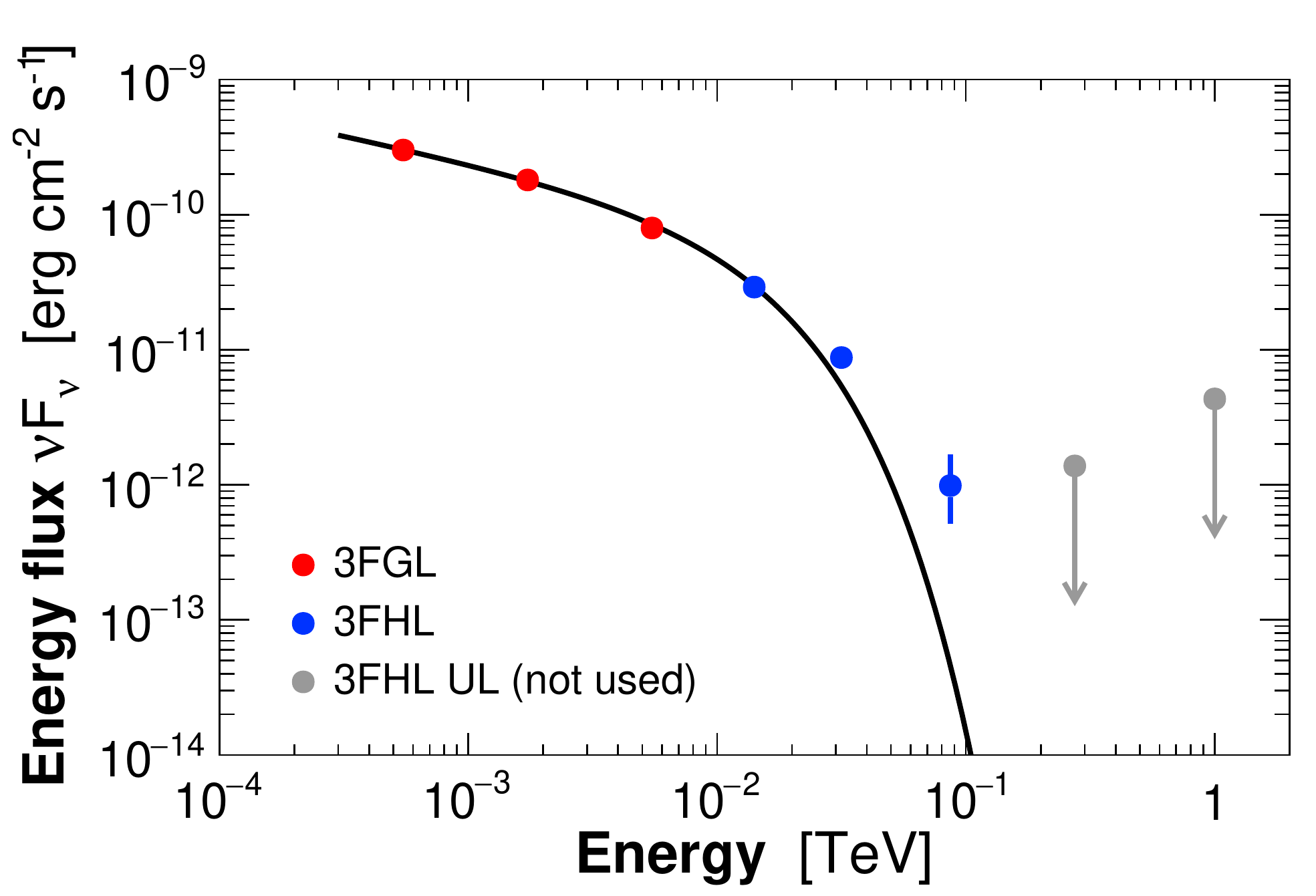}\includegraphics[width=0.45\columnwidth]{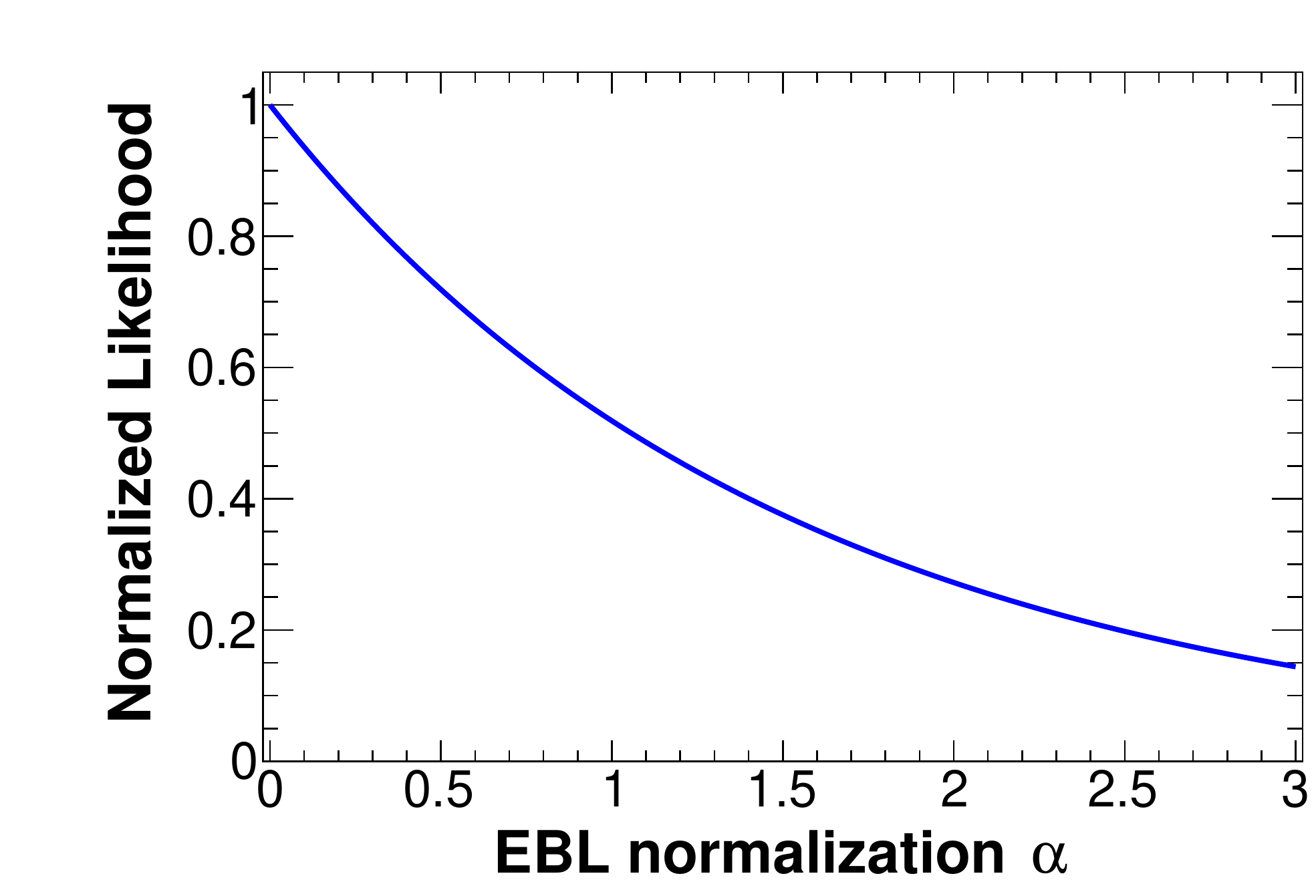}}%
\hfill
\hspace{-3mm}
\subcaptionbox{\label{b}}{\includegraphics[width=0.45\columnwidth]{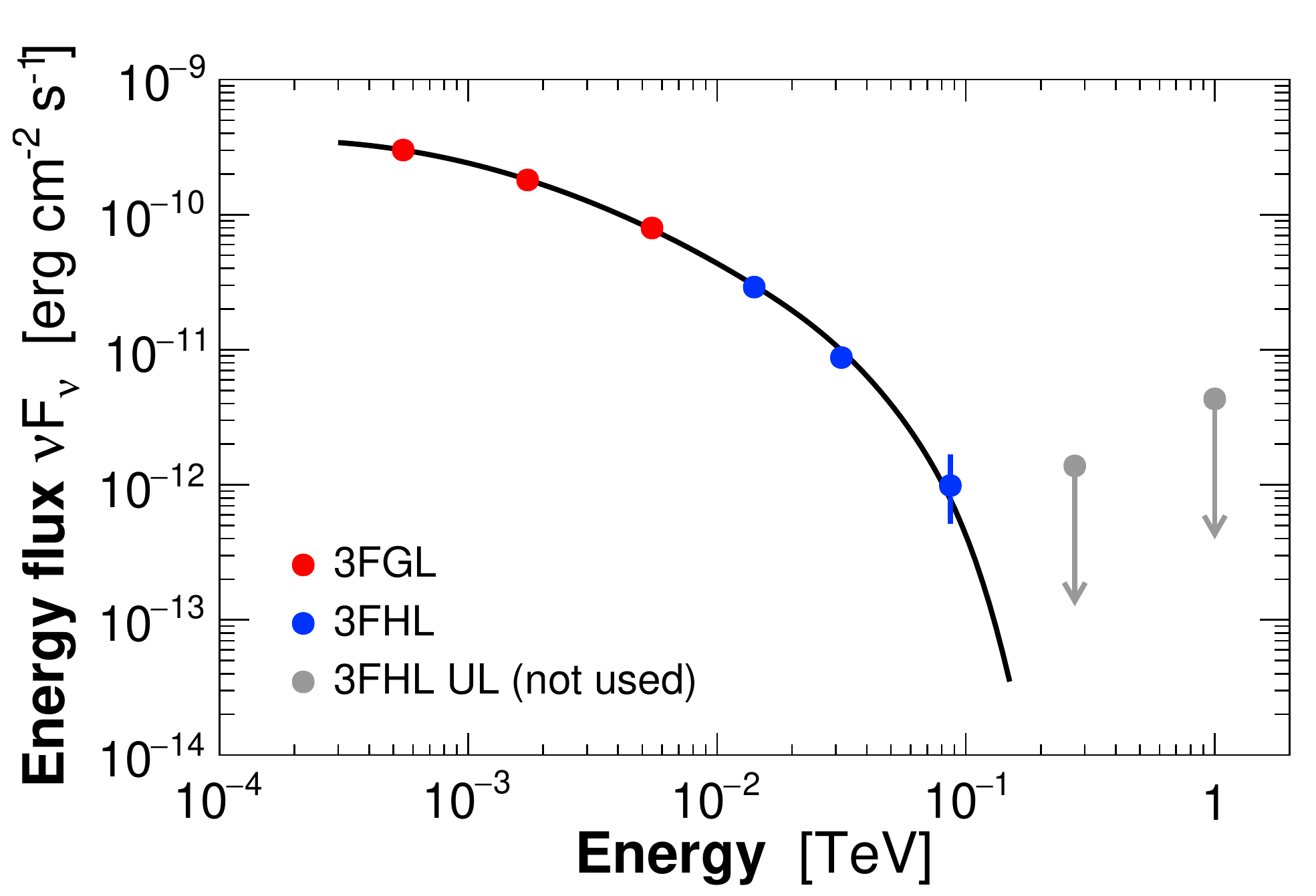}\includegraphics[width=0.45\columnwidth]{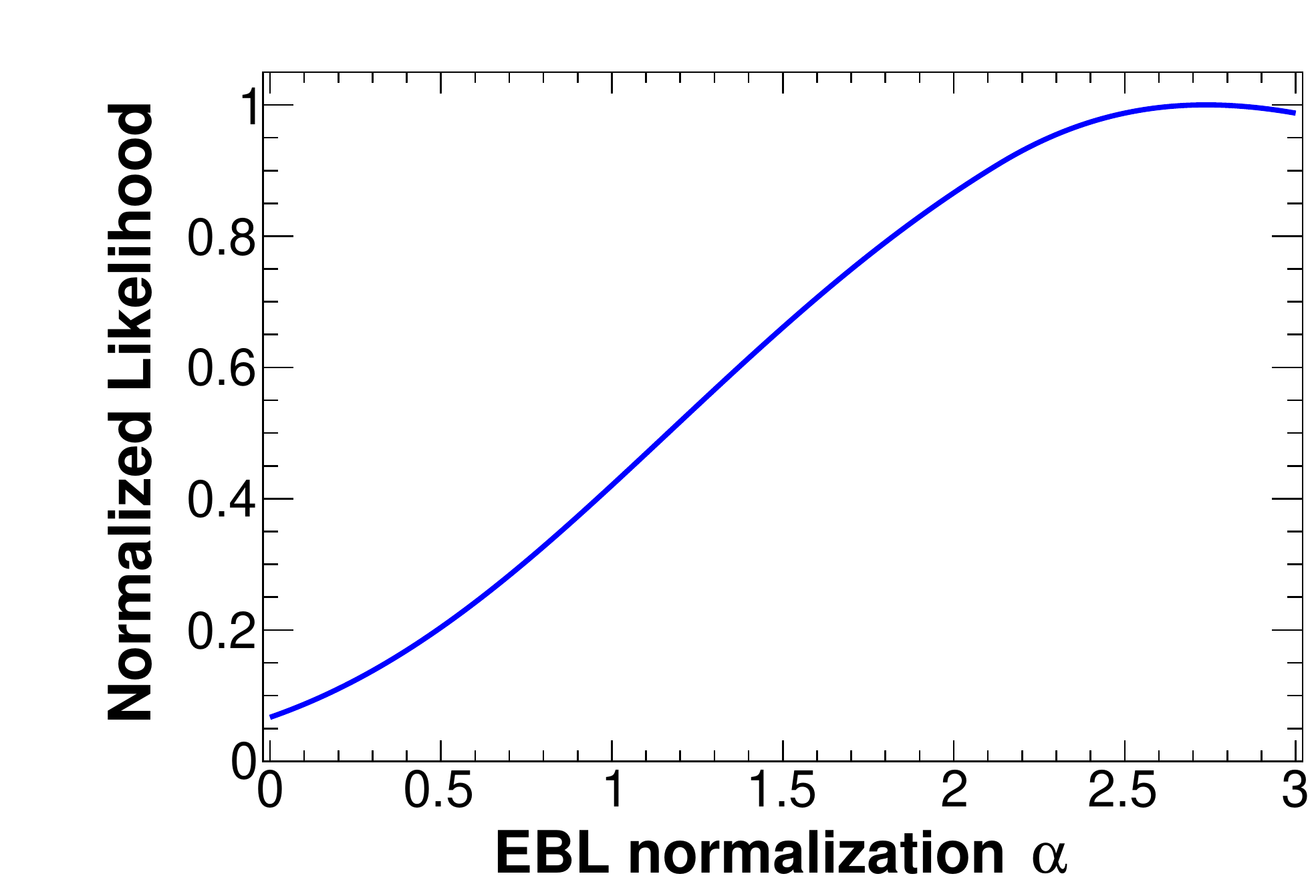}}%
\hfill
\hspace{-3mm}
\subcaptionbox{\label{d}}{\includegraphics[width=0.45\columnwidth]{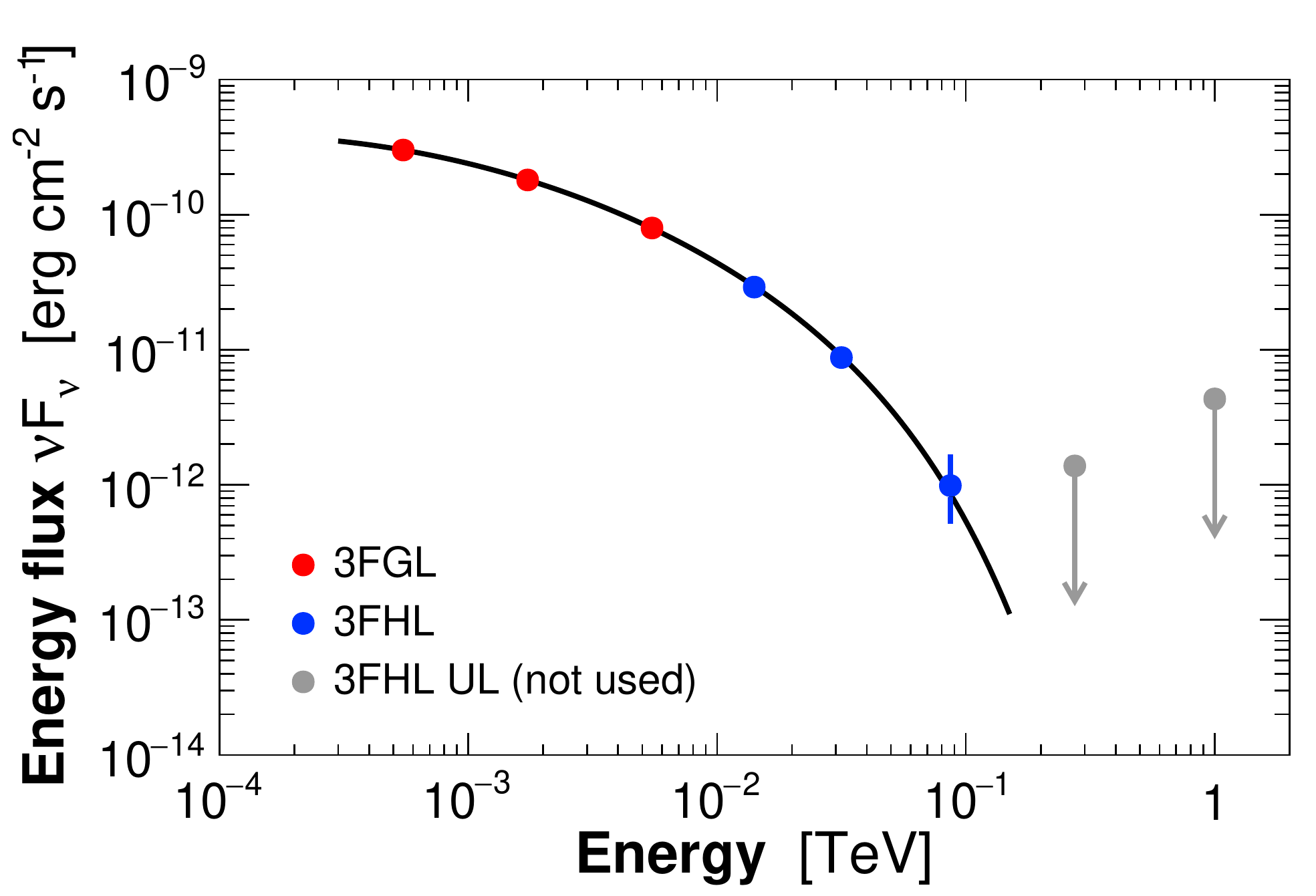}\includegraphics[width=0.45\columnwidth]{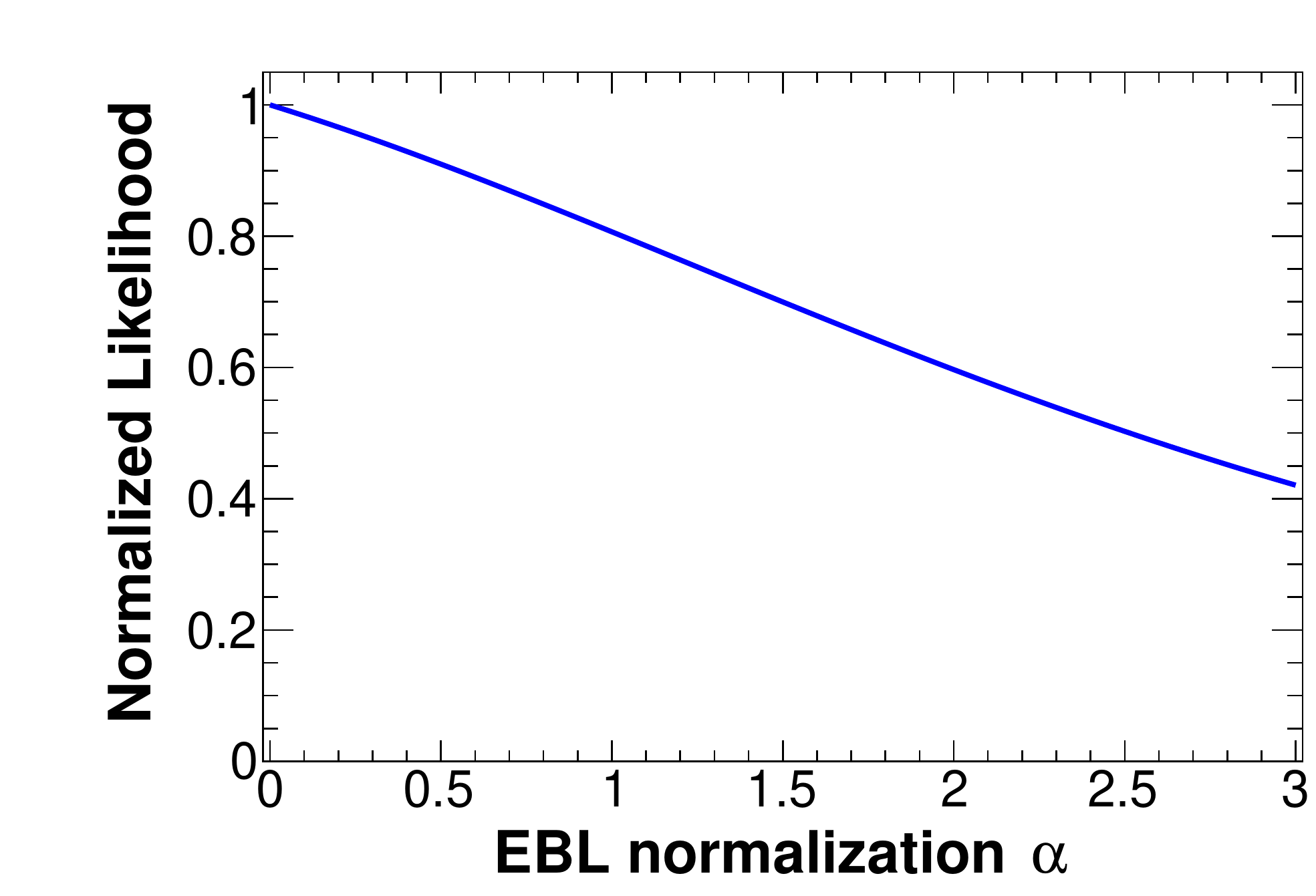}}%
\caption{Left: 3FHL J2253.9+1608 spectrum, including points from the 3FGL catalog (red), and from 3FHL catalog (blue); upper limits (not included in the fit) are shown in gray. The spectral fit is displayed for the best EBL normalization over the range of interest, and has been performed modeling the intrinsic spectrum with a power law (a), an exponential cutoff power law (b), a log-parabola (c), and an exponential cutoff log-parabola (d). Right: Likelihood profile as a function of the EBL normalization, $\alpha$.}
\label{single_source_ex}
\end{figure}

So far, all the methods proposed in the literature that use several $\gamma$-ray emitters to determine the EBL normalization 
make a priori assumptions on the intrinsic spectrum of the sources. The lack of unequivocal arguments to prefer one model over another can lead to biases in the EBL reconstruction.
\cite{Ackermann2012} modeled the intrinsic blazar spectra with a log-parabola function. 
First, the intrinsic spectral parameters are determined by fitting the unabsorbed part of the spectrum. The EBL normalization is then reconstructed through a second fit in the whole range (1-500$\,$GeV) by fixing the curvature to the value obtained in the previous fit.
The choice of a log-parabola parametrization was validated based on the average of the residuals with respect to the best-fit models, on the spectral fit of well-known blazars with coverage in the GeV-TeV band, and on simulations of blazar broad-band spectra.
\cite{Hess2017} also used log-parabola functions to model intrinsic spectra of blazars. A joint fit on each source was performed to obtain both the best spectral parameters and the EBL normalization, using the full energy range covered by the experiment. Finally, all the results were combined together to determine
the final EBL normalization.
In \cite{Armstrong2017}, the authors used two functions to model the unabsorbed part of the spectra: power-law and log-parabola. A likelihood profile was generated as a function of the EBL normalization, and the log-parabola model was chosen if the Test Statistic (TS) between the two models was larger then 16, the same value adopted by the {\it Fermi}-LAT Collaboration to choose a more complex spectral model over a power law in the 3FGL catalog \citep{Acero2015}.
In \cite{Hess2013}, the intrinsic spectral model was chosen according to the highest $\chi^2$ probability of the fit, among power-law, log-parabola, exponential-cutoff power-law, exponential-cutoff log-parabola, and super-exponential-cutoff power-law models. The intrinsic spectral model was chosen individually for each source, and independently from the value of the EBL normalization where the $\chi^2$ probability reaches its maximum.
The same approach was used in \cite{Mazin2017}, but assuming a log-parabola as the simplest function rather than a power law, which could bias the measurement towards low EBL levels.
The same procedure and the same spectral models of \cite{Mazin2007} were used in \cite{Acciari2019}. In this case, the determination of the intrinsic spectra involves a further step. Once a first intrinsic model 
was 
chosen for each source according to the highest {\it p}-value of the fit and independently from the EBL normalization, $\alpha$, a preliminary maximum-likelihood 
fit was
performed using all 32 sources. The intrinsic spectral models 
were then recomputed for all the sources whose models 
were previously
outside the 2$\sigma$ range around the best $\alpha$.
Finally, in \cite{Biteau2015}, the intrinsic spectral models have been selected iteratively in a joint fit together with the EBL normalization. The modeling function is chosen among power-law, log-parabola, exponential-cutoff power-law, and exponential-cutoff log-parabola models. For a given starting set of initial models, in correspondence to the best EBL normalization, a more complex model is selected if preferred at least at the $2\sigma$ level (a threshold commonly adopted in the literature) with respect to a simpler one. This procedure is repeated until convergence of the set of intrinsic models.

The assumptions on the parametrization of the intrinsic curvature may have a strong impact on the EBL reconstruction. A weak disentanglement criterion or an improper choice of the intrinsic models can be responsible for either an over- or underestimation of the EBL normalization. An example is provided in Fig.~\ref{single_source_ex}, where the EBL normalization has been determined from the source 3FHL J2253.9+1608 ($z=0.86$), whose spectrum has been modeled with a power law, an exponential cutoff power law, a log-parabola, and an exponential cutoff log-parabola, respectively. The best spectral parameters are fit together with the EBL normalization. If the intrinsic spectrum is modeled with a power law or an exponential cutoff power law, the $\chi^2$ probability is very low: $6.0\times10^{-88}$ and $6.0\times10^{-7}$, respectively. By using either a log-parabola or an exponential cutoff log-parabola, the fit is better, with a $\chi^2$ probability of 0.33 and 0.19, respectively. Nevertheless, the most-likely EBL normalizations differ a lot: 2.7 for the log-parabola, and zero (i.e., no EBL absorption) for the exponential cutoff log-parabola. The cumulative effect -- over a large number of sources -- due to an improper choice of the intrinsic spectrum, could seriously affect the EBL reconstruction.

\begin{figure*}[ht!]
\centering
\subcaptionbox{\label{Sig_before_corr}}{\includegraphics[width=0.49\columnwidth]{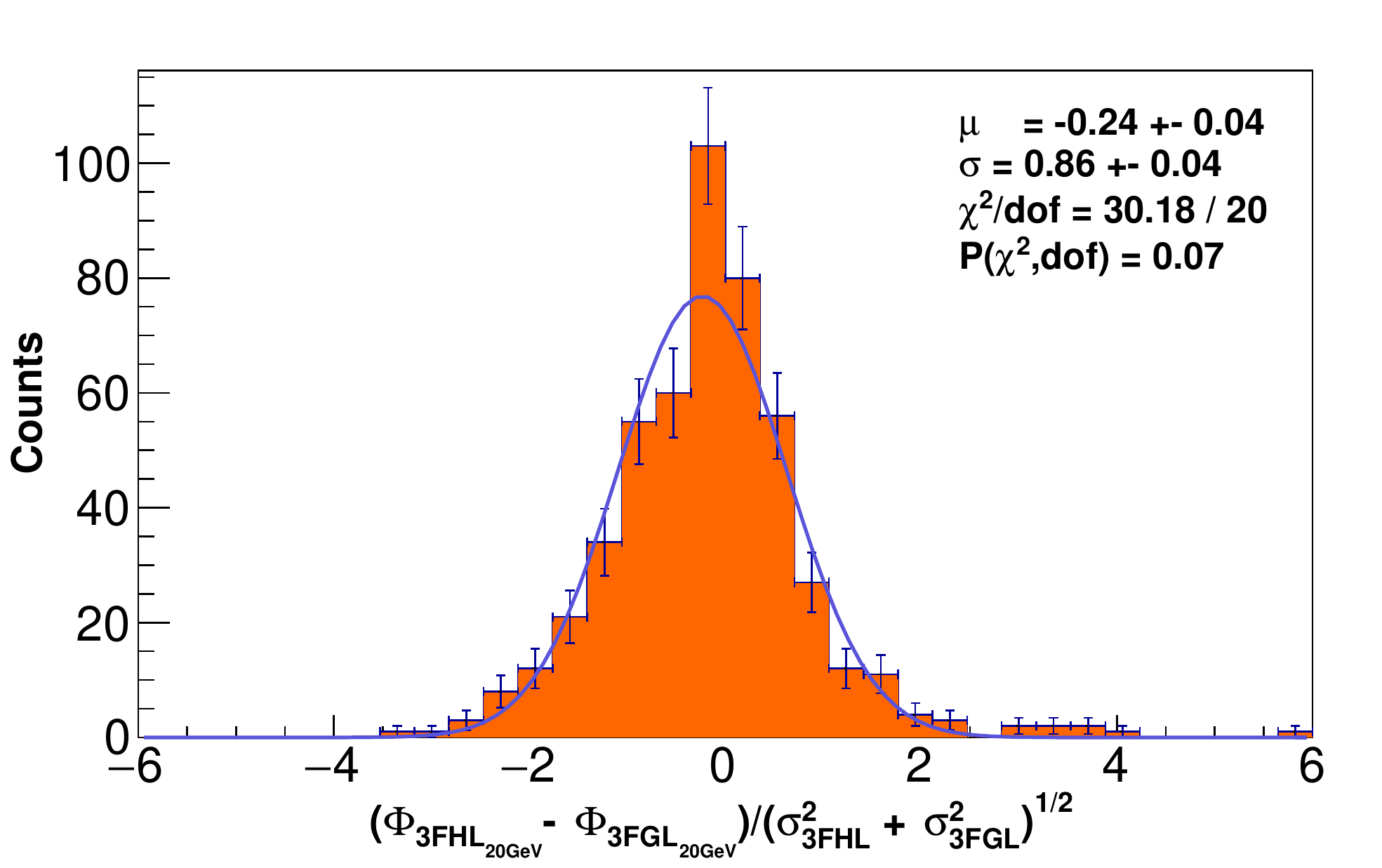}}
\subcaptionbox{\label{Sig_after_corr}}{\includegraphics[width=0.49\columnwidth]{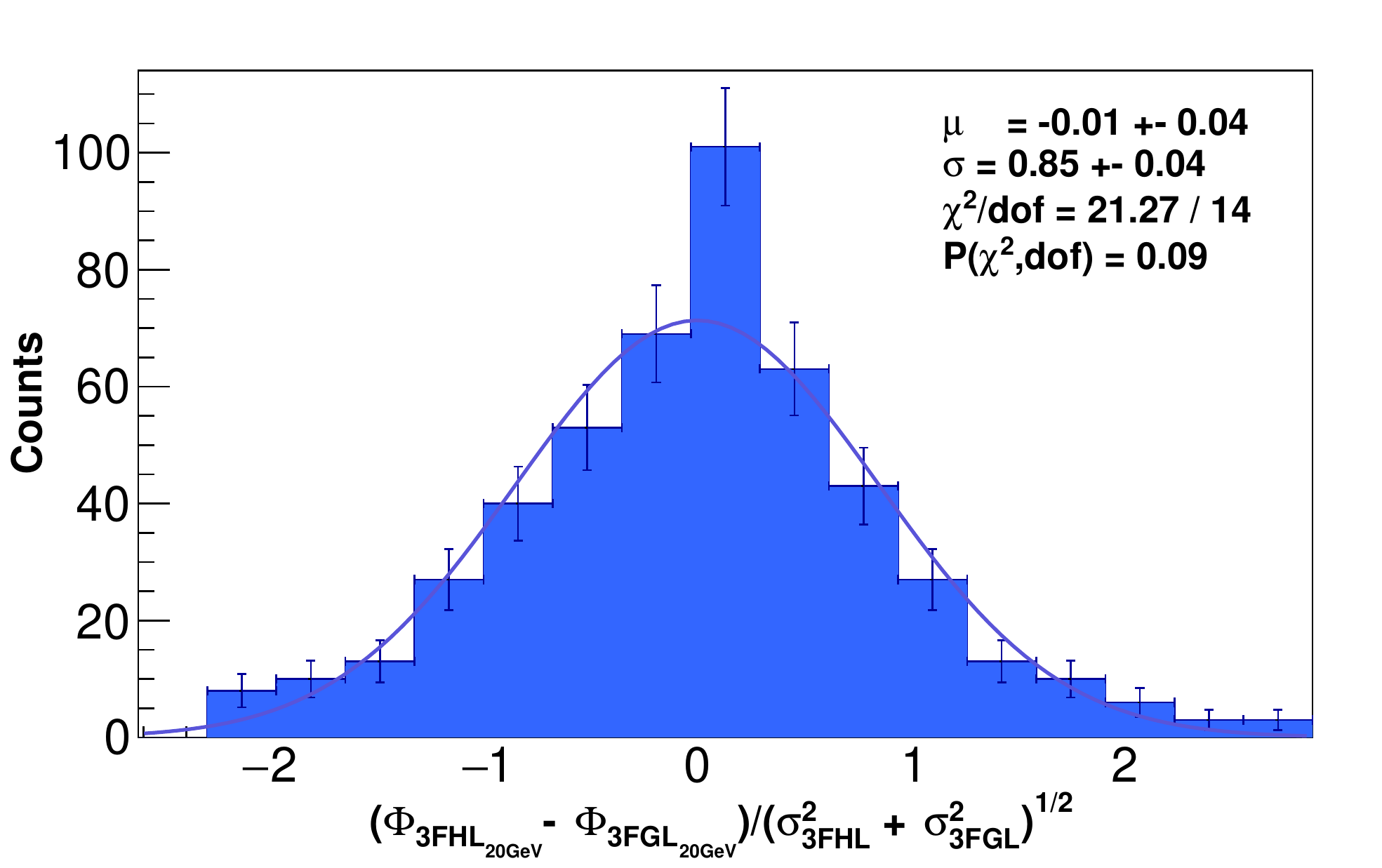}}
\caption{Distribution of the normalized difference between the flux at $20\,$GeV from the 3FGL and 3FHL catalogs before (a) and after (b) the correction (see text for details). A Gaussian function has been fit to both histograms; the mean of the distributions, $\mu$, quantifies the mismatch between the fluxes reported in the two catalogs.}
\label{Histo_distrbution}
\end{figure*}

The main aim of this work is to obtain the EBL normalization using 
an approach that (i) avoids, as much as possible, any assumption on the parametrization of the intrinsic curvature; (ii) justifies the selection criterion among different spectral models; and (iii) determines the best intrinsic spectrum, together with the EBL scaling factor, through 
a rigorous, joint fitting process. This approach, applied to a blazar sample that combines {\it Fermi}-LAT data from two catalogs, is described in Sect. \ref{analysis}, followed by the estimation of the systematic uncertainties in Sect. \ref{results}.

\section{Data sample}
\label{data_sample}

\subsection{Source selection}
\label{source_selection}

To determine the scaling factor for different EBL models, we used AGN spectra both from the Third Catalog of High-Energy 
{\it Fermi}-LAT Sources \citep[3FHL,][]{Fermi3FHL2017} and the Third {\it Fermi}-LAT Source Catalog \citep[3FGL,][]{Acero2015}.
The 3FGL catalog compiles data from the first 4 years of the {\it Fermi}-LAT mission, for a total of 3033 objects detected above 4$\sigma$ significance between $100\,$MeV and $300\,$GeV, while the 3FHL catalog includes data from 7 years of {\it Fermi}-LAT observations, counting 1556 sources detected between 10\,GeV and 2\,TeV. The 3FHL data have been processed with the event-level analysis Pass8 \citep{Atwood2013} that provides significant improvements with respect to the previous reprocessed Pass7 \citep[P7REP,][]{Bregeon2013} used for the 3FGL. With the Pass8 event-level analysis, the sensitivity and the angular resolution have been improved by a factor of two and three, respectively, at the same energies with respect to the previous First Catalog of High-Energy {\it Fermi}-LAT Sources \citep[1FHL,][]{Ackermann2013}.

We select AGNs listed in the 3FHL (version 13\footnote{\url{https://fermi.gsfc.nasa.gov/ssc/data/access/lat/3FHL/}}) with a known redshift, with at least one significant point, and which have a counterpart in the 3FGL catalog. Among 509 sources, 13 were found to be associated with an incorrect or unreliable redshift. The excluded sources are: 3FHL\,J0816.4-1311, 3FHL\,J0449.4-4350, and 3FHL\,J0033.5-1921 whose redshift is a lower-limit estimation \citep{Pita2014}; 3FHL\,J0521.7+2112 because of spectral lines affected by telluric contamination  \citep{Shaw2013}; 3FHL\,J1120.8+4212 \citep{Paiano2017} and 3FHL\,J1436.9+5639 \citep{Shaw2013} for featureless spectra; 3FHL\,J0211.2+1051 \citep{Meisner2010} and 3FHL\,J0650.7+2503 \citep{Rector2003} because of an unreliable photometric redshift; 3FHL\,J0508.0+6737 because of the contamination due to a nearby star \citep{Giovannini2004}; 3FHL\,J1443.9-3908, 3FHL\,J1958.3-3011, 3FHL\,J2324.7-4040, and 3FHL\,J0622.4-2606 for a redshift derived from spectra collected in the 6dF Galaxy Survey \citep{Jones2009} characterized by a low signal-to-noise ratio.
On the other hand, seven other sources without an associated redshift in the 3FHL catalog were found to have a solid measurement: 3FHL\,J0550.5-3115 with $z=0.069$ \citep{Mao2011}; 3FHL\,J1603.8-4903 with $z=0.232$ \citep{Goldoni2016}; 3FHL\,J0237.6-3602 with $z=0.411$ \citep{Pita2014}; 3FHL\,J1442.5-4621 with $z=0.103$ from the 6dF Galaxy Survey \citep{Jones2009}; 3FHL\,J1410.5+1438 with $z=0.144$ and 3FHL\,J0022.0+0006 with $z=0.306$ from the Sloan Digital Sky Survey DR14 \citep{Abolfathi2018}; and finally 3FHL\,J0338.9-2848 with $z=0.251$ from \cite{Halpern1997}. However, only three of them (3FHL\,J0550.5-3115, 3FHL\,J1603.8-4903, and 3FHL\,J0237.6-3602), with at least one significant point in their 3FHL spectrum,
have been included in the sample.

\subsection{Source spectra} 
\label{spectral_points}

In order to assess the compatibility between the fluxes reported in the 3FHL and 3FGL catalogs which were analyzed with different instrument response functions, we inspected the distribution of $(\Phi_{3FHL} -  \Phi_{3FGL})/(\sigma_{3FHL}^2 + \sigma_{3FGL}^2)^{1/2}$, where $\Phi_{3FGL}$ and $\Phi_{3FHL}$ are the fluxes at 20\,GeV --- the energy at the middle of the energy-range overlap between the two catalogs --- calculated from the spectral parameters reported in the 3FGL catalog and the 3FHL catalog, respectively. $\sigma_{3FGL}^2$ and $\sigma_{3FHL}^2$ are the related errors obtained by propagating the errors on the fit parameters. The histogram shown in Fig.~\ref{Sig_before_corr} points out both the presence of outliers and a significant shift from zero of the mean of the distribution ($-0.24 \pm 0.04$).

To determine the offset between the fluxes at 20\,GeV and the ensuing overall scaling factor to apply to the 3FGL spectral points, the following approach was adopted: (i) the mean of the distribution and its associated error computed through a Gaussian fit --- including all the sources --- was calculated; (ii) sources that show a deviation from the mean larger than $3 \sigma$ were excluded; (iii) the quantity $r = \log (\Phi_{3FHL} / \Phi_{3FGL})$ was calculated, and the 3FGL spectral points were corrected for a factor $e^{r}$; (iv) the procedure from point (i) was repeated until no more outliers were found.
The procedure converged at the first iteration, and the final scaling factor, $r$, is $-0.10 \pm 0.03$, which corresponds to a scaling in flux of $0.90 \pm 0.03$.
Among the whole sample, nine sources were identified as outliers and removed from the sample: 3FHL\,J0510.0+1800 (PKS\,0507+17), 3FHL\,J0958.7+6533 (QSO\,B0954+65), 3FHL\,J1104.4+3812 (Mrk\,421), 3FHL\,J1230.2+2517 (ON\,246), 3FHL\,J1415.6+4830, 3FHL\,J1427.9-4206 (PKS\,1424-418), 3FHL\,J1443.9+2502 (PKS\,1441+25), 3FHL\,J1522.6-2731, and 3FHL\,J1728.3+5013 (QSO\,B1727+502).

A possible explanation concerning the flux mismatch between the two catalogs lies in the high variability of the sources: strong flares can contribute to slightly modify the average flux level over two different time-spans. In order to investigate this hypothesis, the distribution of the 3FGL variability index was inspected. Six out of nine sources show a variability index above a threshold of 72.44 (indicating a probability of 1\% of being a steady source): 3FHL\,J0510.0+1800 (114.8), 3FHL\,J0958.7+6533 (210.7), 3FHL\,J1104.4+3812 (755.1), 3FHL\,J1230.2+2517 (95.3), 3FHL\,J1427.9-4206 (3146.8), and 3FHL\,J1522.6-2731 (182.9). The three remaining outliers are characterized by a variability index below the threshold: 
3FHL\,J1415.6+4830 (53.9), 3FHL\,J1443.9+2502 (48.0), and 3FHL\,J1728.3+5013 (54.1). This indicates that variability may play a role, but it might not be the only factor responsible for the flux mismatch.

Figure~\ref{Sig_after_corr} shows the distribution obtained after correction for a scaling factor $r =-0.10$. The distribution is well centered on zero, but shows a spread slightly smaller than one. This small discrepancy with a standard normal distribution is expected since we are dealing with nonindependent datasets containing double counting of data from catalogs characterized by different statistics.

To investigate the effect of the variability, we split the sample (including the nine outlier sources mentioned above)
into steady and variable sources, according to the estimators from both the 3FGL and the 3FHL catalogs. In particular, we classified as steady sources those characterized by a variability index (3FGL) $<71.5$ and a number of Bayesian blocks (3FHL) $=1$. This cut results in two subsamples, steady and variable, containing 249 and 250 sources, with associated scaling factors of $-0.02 \pm 0.03$ and $-0.22 \pm 0.03$, respectively.
Such a difference could derive from the fact that the 3FHL catalog contains 3 years more data than the 3FGL. Since steady sources should not show detectable variability, no measurable change is expected in their flux over long periods. Variable sources on the contrary might have shown variability over the three extra years covered by the 3FHL. Furthermore, we note that 3FHL data have been processed with the event-level analysis Pass8, while Pass7 was used for the 3FGL. Given that no firm conclusion can be drawn as to the origin of the discrepant scaling factor, we use $r = - 0.1 \pm 0.1$ and treat the uncertainty as a systematic effect (see Sect.~\ref{results}).

To summarize, starting from a list of 509 sources, 13 were excluded because of an incorrect or unreliable redshift, 3 sources were added that have a solid redshift measurement and at least one significant point in the 3FHL spectrum, and 9 outlying sources with a large difference between the 3FGL and 3FHL flux were also excluded.
The final sample contains 490 sources with a redshift between 0.003 and 2.534. To be conservative, the first point of the 3FGL catalog (at $173\,$MeV) was excluded from the analysis because of the low {\it Fermi}-LAT acceptance.\footnote{\url{https://www.slac.stanford.edu/exp/glast/groups/canda/archive/p7rep\_v15/lat\_Performance.htm}}
Both catalogs report the flux measured at $31.6\,$GeV. The 3FHL point at $31.6\,$GeV was included in the fit because of a much larger statistic.
Finally, the upper limits (ULs) were not considered in the fitting procedure due to the lack of a robust way of treating them.

\section{Analysis method}
\label{analysis}

\subsection{Extragalactic background light  absorption} 

\begin{figure*}[t!]
\hspace{-2mm}
\includegraphics[scale=0.45]{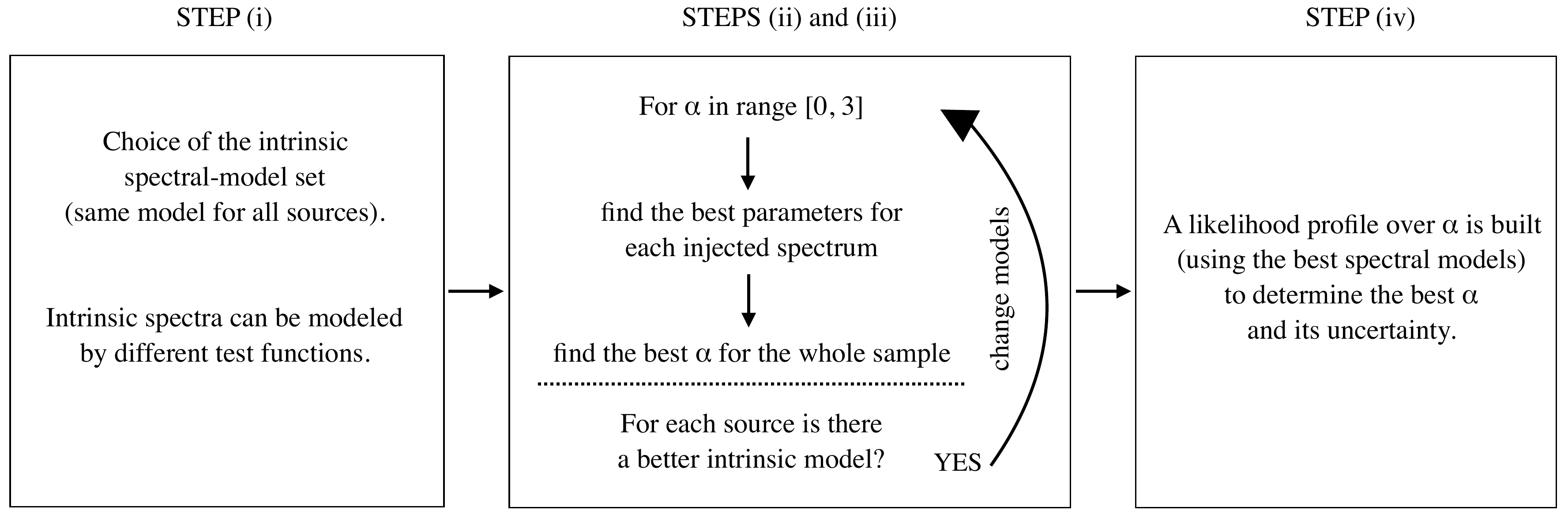}
\caption{Diagram illustrating the steps of the fitting procedure.}
\label{diagram_fit}
\end{figure*}

The process responsible for the $\gamma$-ray opacity is the electron--positron pair production that occurs when $\gamma$-rays interact with photons of the EBL. 

\noindent The effect of the $\gamma$-ray attenuation is encoded in the EBL optical depth:

\begin{equation}
\label{optical depth}
\begin{split}
\tau{(E_0, z_0)} = & \int^{z_0}_0 dz \frac{\partial L}{\partial z}(z) \int^{\infty}_0 d\epsilon \frac{\partial n}{\partial \epsilon}(\epsilon , z) \\
                   & \int^{-1}_1 d\cos \theta \frac{1- \cos \theta}{2} \sigma_{\gamma \gamma} E_0(1+z), \epsilon,\cos \theta)] {\text  ,}
\end{split}
\end{equation}

\noindent that is given by multiplying the number density of the background photon field, $n$, with the process cross section, $\sigma_{\gamma \gamma}$, and then by integrating over the distance, the scattering angle, and the energy of the background photons in the comoving frame. 
The cross-section reaches its maximum when the energy of EBL photons is $\sim$1 \, ${\rm eV} \times (E/500 \, {\rm GeV} )^{-1}$, which means that high-energy $\gamma$-rays mostly interact with COB photons, while VHE photons interact with CIB ones. Finally, ultra-high-energy photons (UHE > 1\,PeV) interact mainly with photons of the CMB.

The effect of the EBL on $\gamma$-ray emission of blazars leaves an imprint on their spectra. In particular, the spectrum is attenuated by a factor:

\begin{equation}
\label{observed_spectrum}
\Phi_{\rm obs} = e^{-\tau(E_0,z_0)} \, \Phi_{\rm intr} {\text  ,}
\end{equation}

\noindent where $\Phi_{\rm obs}$ and $\Phi_{\rm intr}$ are the observed and intrinsic spectrum, respectively, and $\tau(E_0,z_0)$ is the EBL optical depth. The total amount of absorption depends on the energy of the $\gamma$-ray, the source distance, and the density of the EBL photon field. Since direct measurements of the EBL are not easy to obtain for the reasons explained in Sect.~\ref{introduction}, many EBL models have been built to estimate the EBL optical depth as a function of redshift, and to derive the ensuing absorption in spectra of high-energy sources.

In this paper, we use the EBL models of \cite{Franceschini2008} (FR08 hereafter), \cite{Dominguez2011} (DOM11), \cite{Gilmore2012} (GIL12), and \cite{Franceschini2017, Franceschini2018} (FR17).

\subsection{Likelihood analysis} 
\label{like_analysis}\label{Sec:likeli}

An EBL scaling factor, $\alpha$, is introduced \citep[\textit{e.g.},][]{Ackermann2012, Hess2013, Biteau2015}, and Eq.~\ref{observed_spectrum} can be rewritten as:

\begin{equation}
\label{observed_spectrum_alpha}
\Phi_{\rm obs} = e^{- \alpha \tau(E_0,z_0)} \, \Phi_{\rm intr} {\text  ,}
\end{equation}

\noindent where the parameter $\alpha$ indicates the agreement between a given EBL model and the $\gamma$-ray observations: $\alpha = 0$ corresponds to the absence of absorption, while $\alpha = 1$ implies that the model predictions are in perfect agreement with the $\gamma$-ray data. The main goal of this work is to derive the scaling factor for different state-of-the-art EBL models as a function of redshift to test the model predictions against $\gamma$-ray data. By combining several blazar spectra, it is possible to derive the mean deviation between the observed and the intrinsic spectra, and hence the EBL optical depth, $\tau(E_0,z_0)$.

We performed a joint fit of the EBL scaling factor, $\alpha$, and of the intrinsic spectral parameters of all the sources. The fitting procedure consists in four main steps:
(i) A common functional form, one of four different spectral shapes given in Eqs.~\ref{eq_pwl}--\ref{eq_elp}, is chosen to model the intrinsic spectrum, $\Phi_{\rm intr}$, of all the sources.
(ii) For $\alpha$ from 0 to 3 (with step of 0.02), the best-fit spectral parameters of each source are determined, together with the best $\alpha$ according to the cumulative $\chi^2$ of the set of individual fits.
(iii) Alternative $\gamma$-ray spectral models are considered for each source, fixing $\alpha$ to its best value obtained in ii).  All of the other possible models that have at least one degree of freedom in the spectral fit are tested for each source.
If another model is preferred by at least $2\sigma$ (see arguments in the following section), it is chosen as the new model for that source. If more than one model is preferred, the simplest model (in the case of a different number of degrees of freedom) or the most preferred one (in case of the same number of degrees of freedom) is selected.
This iterative selection (ii-iii) keeps going on until the convergence on the model set is reached. This prevents the results from being biased by an inappropriate choice of intrinsic $\gamma$-ray spectral models.
(iv) Finally, the likelihood profile is computed as a function of $\alpha$ in order to find the statistical uncertainty associated with the best scaling factor. The full fitting procedure is illustrated in the diagram of Fig. \ref{diagram_fit}.

\subsection{\texorpdfstring{$\gamma$-}-ray spectral model selection} 
\label{model_selection}

Equation~\ref{observed_spectrum} points out that an accurate estimate of the EBL depends crucially on the assumptions on the intrinsic spectral shape, especially if we consider that the emission processes of blazars are still not fully understood. In other words, it may seem difficult at first to disentangle intrinsic curvature from interaction with EBL photons.

In order to avoid bias in the results due to an inappropriate choice of intrinsic spectral models, various functions have been tested for each source: power law (PWL, hereafter)

\begin{equation}
\label{eq_pwl}
\Phi_{\rm PWL}(E) = \Phi_0 \left( \frac{E}{E_0} \right)^{-a} {\text  ,}
\end{equation}

\noindent exponential cutoff power law (EPWL),

\begin{equation}
\Phi_{\rm EPWL}(E) = \Phi_0 \left( \frac{E}{E_0} \right)^{-a} e^{-E/E_{\rm cut}} {\text  ,}
\end{equation}

\noindent log-parabola (LP),

\begin{equation}
\Phi_{\rm LP}(E) = \Phi_0 \left( \frac{E}{E_0} \right)^{-a -b \ln(E/E_0)} {\text  ,}
\end{equation}

\noindent and exponential cutoff log-parabola (ELP),

\begin{equation}
\label{eq_elp}
\Phi_{\rm ELP}(E) = \Phi_0 \left( \frac{E}{E_0} \right)^{-a -b \ln(E/E_0)} e^{-E/E_{\rm cut}} {\text  ,}
\end{equation}

\noindent where $\Phi_0$ is the flux normalization, $E_0$ is a reference energy (set as $\sqrt{E_{\rm min} \, E_{\rm max}}$, where $E_{\rm min}$ and $E_{\rm max}$ are the minimum and the maximum energy of the \textit{Fermi}-LAT spectrum, respectively), $a$ is the spectral index, $b$ is the curvature parameter, and $E_{\rm cut}$ is the energy corresponding to the cutoff energy.

\begin{figure}[t!]
\includegraphics[width=8cm]{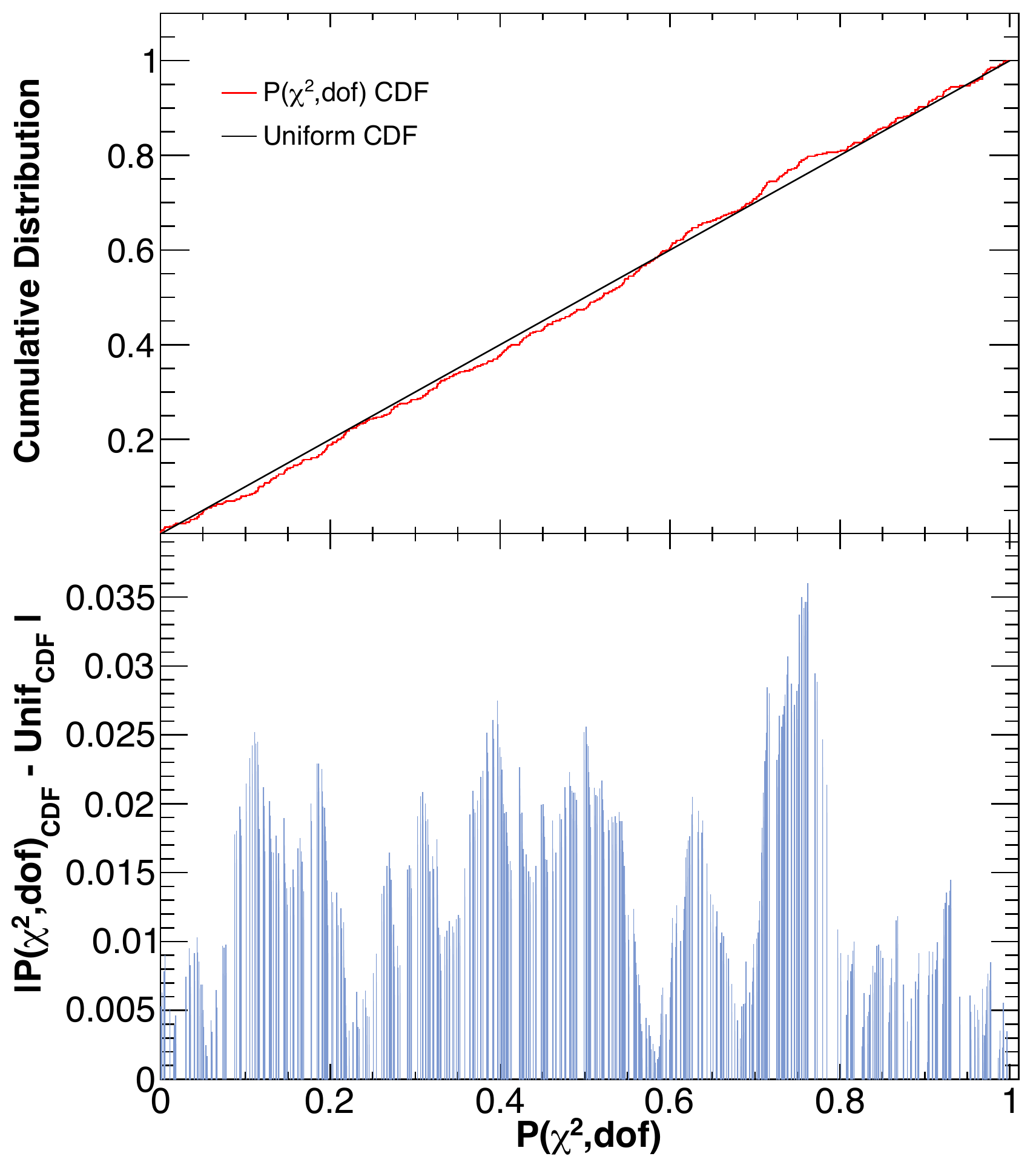} 
\caption{Example of the Kolmogorov-Smirnov statistic obtained for the whole sample, by injecting LP functions as intrinsic spectral models and for $\sigma=2$ (see text for details). Top:  Comparison between the CDF of the $\chi^2$ probabilities resulting from the fit and the CDF of a uniform distribution. Bottom: Bar chart of the distance between the two CDFs as a function of each measured $\chi^2$ probability.}
\label{KS_prob}
\end{figure}

The best spectral model for each source is computed iteratively in a procedure explained in Sect.~\ref{like_analysis}. Since the approach we follow is not based on a specific assumption on the intrinsic spectral models, a criterion to discriminate among different models is needed. The switch occurs if a model is preferred at a certain $\sigma$ level with respect to another one:\footnote{In the selection process the priority is given to simpler models, {\it i.e.}, simpler models are compared to more and more complex ones.}

\begin{equation}
\sigma = \sqrt{2} \, \text{erfc}^{-1} \left[ P(\Delta \chi^2, \Delta \text{dof}) \right] {\text  ,}
\end{equation}

\noindent where erfc$^{-1}$ is the inverse complementary error function, $\Delta \chi^2$ is the difference in the $\chi^2$ obtained for the two comparing models, and $\Delta$dof is the difference of the number of parameters of the two models. We note that the PWL-LP-ELP and PWL-EPWL-ELP models presented above are nested, enabling a straightforward estimation of the significance. In cases where a choice between an LP and EPWL model is needed, the model with the largest significance with respect to a PWL is chosen.

\begin{figure}[t!]
\includegraphics[width=9cm]{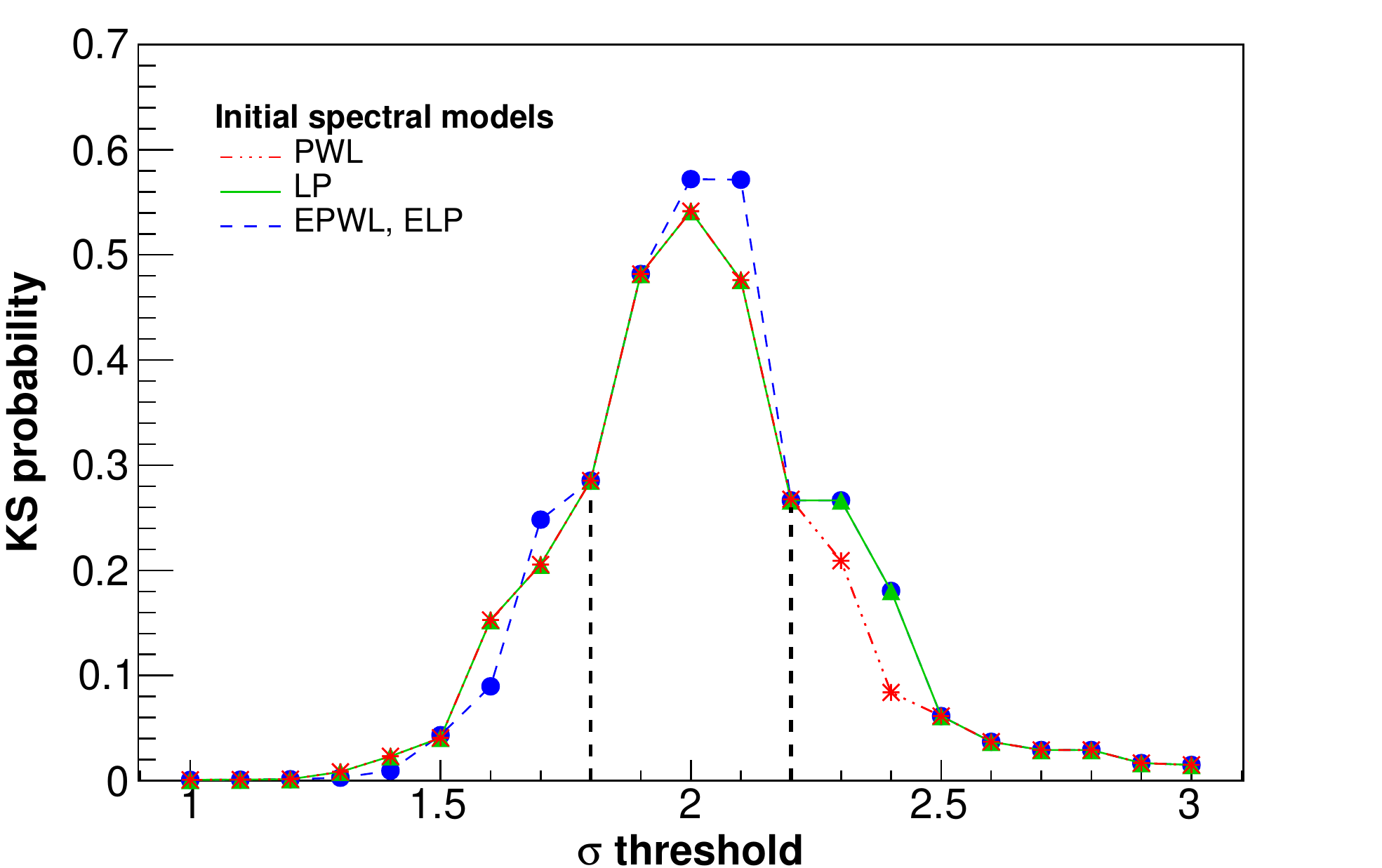}
\caption{Kolmogorov-Smirnov probability as a function of the $\sigma$ threshold used to switch to a more complex intrinsic spectral model (results obtained starting from EPWL and ELP models are coincident). The results were obtained by using the EBL model of \cite{Gilmore2012} and different starting spectral models, as discussed in Sect.~\ref{Sec:likeli}. Dotted vertical lines at 1.8 and 2.2 correspond to $\sigma$ values used to estimate the systematic uncertainties due to the model selection (see text for details).}
\label{all_KS_prob}
\end{figure}

To determine the best $\sigma$ threshold, we checked the compatibility between the observed $\chi^2$-probability distribution over the sample of sources and a uniform distribution through a Kolmogorov-Smirnov (KS) test (an example is shown in Fig. \ref{KS_prob}). For several $\sigma$ values in the range $1 \le \sigma \le 3$, the KS probability of obtaining a deviation, $D$, between the observed and ideal distribution of the $\chi^2$ probability distribution of the individual spectra, that is larger than the observed one (in case the null hypothesis is true, {\it i.e.}, the two distributions are the same) has been calculated as \citep{Num_Rec_2007}:

\begin{equation}
P(D<{\rm observed}) = P_{KS} \left[ \left( \sqrt{N_e} + 0.12 +\frac{0.11}{\sqrt{N_e}} \right) D \right] {\text  ,}
\end{equation}

\noindent where the CDF, $P_{KS}(z)$ (defined for positive $z$), is defined by the series:

\begin{equation}
P_{KS}(z) = 2 \sum_{j=1}^{\infty} (-1)^{j-1} e^{-2j^2z^2} {\text  .}
\end{equation}

The results obtained starting from different sets of initial spectral models are reported in Fig. \ref{all_KS_prob}. Although the initialization is done with different spectral models, the final selected ones can end up being the same ({\it i.e.} coincident points in Fig. \ref{all_KS_prob}), and this comes in favor of the robustness of the model selection.
For low ({\it e.g.}, $1\sigma$) and high $\sigma$ values ({\it e.g.}, $3\sigma$), the KS probability is low, which means that the $\chi^2$-probability distribution is far from being compatible with a uniform distribution, which is the distribution expected in case of a healthy fit procedure. Conversely, higher values of the KS probability indicate a compatibility between the $\chi^2$ probability and a uniform distribution.\footnote{We investigated the presence of outliers at a  $\chi^2$ probability close to zero and one. We find three spectra with a $\chi^2$ probability lower than 0.1\%, which is expected with a probability of 0.1\% for a sample of 490 objects. We verified that excluding these three sources from the analysis does not impact our results. Similar conclusions are drawn for each of the subsamples studied in this paper.}

A Gaussian was fitted to each set between 1.4 and 2.6, and all of them lead to a mean $\sigma$ threshold of 2.0 and a standard deviation of 0.2. Therefore, the adopted threshold for the model selection is $\sigma=2.0$ (corresponding to the highest KS probability), and the standard deviation is used to estimate the systematic uncertainty related to the model selection.

\afterpage{
\renewcommand{\arraystretch}{1.7}
\begin{table*}[h!]
\centering
    \caption{The EBL scaling factors $\alpha$ and related uncertainties obtained for different EBL models.}
    \label{results_table1}
    \begin{tabular}{c c c c c c c }
    \hline
    \hline
      $z$    &   EBL model &$\alpha \pm$    & PWL  & LP   & EPWL & ELP  \\
             &             & stat. err      & [\%] & [\%] & [\%] & [\%] \\   
      \hline
      \hline
      all       &  FR17&  $1.04_{{-0.10}_{\rm stat}{-0.28}_{\rm syst}}^{{+0.10}_{\rm stat}{+0.31}_{\rm syst}}$& 74.5 & 15.9 & 9.6 & 0.0 \\ 
            &  GIL12& $1.05_{{-0.11}_{\rm stat}{-0.24}_{\rm syst}}^{{+0.12}_{\rm stat}{+0.32}_{\rm syst}}$& 74.2 & 15.1 & 10.7 & 0.0 \\ 
            &  DOM11& $1.08_{{-0.13}_{\rm stat}{-0.35}_{\rm syst}}^{{+0.13}_{\rm stat}{+0.40}_{\rm syst}}$& 73.9 & 14.9 & 11.2 & 0.0 \\ 
            &  FR08&  $1.13_{{-0.12}_{\rm stat}{-0.17}_{\rm syst}}^{{+0.12}_{\rm stat}{+0.21}_{\rm syst}}$& 74.7 & 14.7 & 10.6 & 0.0 \\ 
      \hline
      \hline
      FSRQs                                       &  FR17&  $1.10_{{-0.20}_{\rm stat}{-0.59}_{\rm syst}}^{{+0.20}_{\rm stat}{+0.16}_{\rm syst}}$ & 55.4 & 29.9 & 14.6 & 0.0 \\ 
      (all)                               &  GIL12& $1.05_{{-0.18}_{\rm stat}{-0.53}_{\rm syst}}^{{+0.20}_{\rm stat}{+0.16}_{\rm syst}}$ & 55.4 & 29.9 & 14.6 & 0.0 \\
      $0.097 \le z \le 2.534$ &  DOM11& $1.17_{{-0.23}_{\rm stat}{-0.62}_{\rm syst}}^{{+0.24}_{\rm stat}+{0.17}_{\rm syst}}$ & 54.8 & 29.3 & 15.9 & 0.0 \\ 
                                          &  FR08&  $1.18_{{-0.19}_{\rm stat}{-0.23}_{\rm syst}}^{{+0.20}_{\rm stat}{+0.17}_{\rm syst}}$ & 56.7 & 28.7 & 14.6 & 0.0 \\
      \hline
      BL~Lacs                              &  FR17&  $1.09_{{-0.12}_{\rm stat}{-0.43}_{\rm syst}}^{{+0.12}_{\rm stat}{+0.18}_{\rm syst}}$  & 83.4 & 9.4 & 7.2 & 0.0 \\
      (all)                                &  GIL12& $1.25_{{-0.16}_{\rm stat}{-0.39}_{\rm syst}}^{{+0.16}_{\rm stat}{+0.43}_{\rm syst}}$  & 83.3 & 8.6 & 8.1 & 0.0 \\
      $0.032 \le z \le 2.471$  &  DOM11& $1.22_{{-0.15}_{\rm stat}{-0.62}_{\rm syst}}^{{+0.16}_{\rm stat}{+0.17}_{\rm syst}}$  & 82.9 & 7.6 & 8.7 & 0.0 \\       
                                           &  FR08&  $1.20_{{-0.15}_{\rm stat}{-0.43}_{\rm syst}}^{{+0.16}_{\rm stat}{+0.51}_{\rm syst}}$  & 83.3 & 8.2 & 8.5 & 0.0 \\
      \hline
      \hline
      $0<z<0.21$ &  FR17&  $1.75_{{-0.33}_{\rm stat}{-0.44}_{\rm syst}}^{{+0.34}_{\rm stat}{+0.68}_{\rm syst}}$& 85.8 & 8.5 & 5.7 & 0.0 \\ 
                 &  GIL12& $1.92_{{-0.43}_{\rm stat}{-0.51}_{\rm syst}}^{{+0.45}_{\rm stat}{+0.66}_{\rm syst}}$& 85.8 & 8.1 & 6.1 & 0.0 \\ 
                 &  DOM11& $1.73_{{-0.39}_{\rm stat}{-0.47}_{\rm syst}}^{{+0.44}_{\rm stat}{+0.67}_{\rm syst}}$& 85.4 & 8.5 & 6.1 & 0.0 \\ 
                 &  FR08&  $1.64_{{-0.37}_{\rm stat}{-0.42}_{\rm syst}}^{{+0.41}_{\rm stat}{+0.60}_{\rm syst}}$& 85.4 & 8.5 & 6.1 & 0.0 \\ 
      \hline   
      $0.21\le z < 0.456$       &  FR17&  $0.51_{{-0.15}_{\rm stat}{-0.17}_{\rm syst}}^{{+0.16}_{\rm stat}{+0.23}_{\rm syst}}$& 85.4 & 4.9 & 9.8 & 0.0 \\ 
                                        &  GIL12& $0.59_{{-0.21}_{\rm stat}{-0.18}_{\rm syst}}^{{+0.24}_{\rm stat}{+0.19}_{\rm syst}}$& 84.6 & 4.9 & 10.6 & 0.0 \\ 
                                        &  DOM11& $0.54_{{-0.19}_{\rm stat}{-0.18}_{\rm syst}}^{{+0.21}_{\rm stat}{+0.18}_{\rm syst}}$& 84.6 & 4.9 & 10.6 & 0.0 \\ 
                                        &  FR08&  $0.53_{{-0.19}_{\rm stat}{-0.17}_{\rm syst}}^{{+0.21}_{\rm stat}{+0.18}_{\rm syst}}$& 84.6 & 4.9 & 10.6 & 0.0 \\
      \hline
      $0.456 \le z < 0.944$ &  FR17&  $1.00_{{-0.32}_{\rm stat}{-1.00}_{\rm syst}}^{{+0.37}_{\rm stat}{+1.18}_{\rm syst}}$& 65.9 & 23.6 & 10.2 & 0.4 \\
                                        &  GIL12& $1.00_{{-0.30}_{\rm stat}{-0.86}_{\rm syst}}^{{+0.33}_{\rm stat}{+1.4}_{\rm syst}}$& 66.1 & 24.2 & 9.3 & 0.4 \\ 
                                        &  DOM11& $0.97_{{-0.31}_{\rm stat}{-0.86}_{\rm syst}}^{{+0.36}_{\rm stat}{+1.11}_{\rm syst}}$& 65.9 & 23.6 & 10.2 & 0.4 \\
                                        &  FR08&  $1.01_{{-0.33}_{\rm stat}{-0.91}_{\rm syst}}^{{+0.40}_{\rm stat}{+1.24}_{\rm syst}}$& 66.1 & 23.6 & 10.0 & 0.4 \\ 
      \hline
      $0.944 \le z \le 2.534$ &  FR17&  $1.30_{{-0.16}_{\rm stat}{-0.27}_{\rm syst}}^{{+0.17}_{\rm stat}{+0.23}_{\rm syst}}$& 63.2 & 22.7 & 14.0 & 0.0 \\
                                          &  GIL12& $1.13_{{-0.14}_{\rm stat}{-0.25}_{\rm syst}}^{{+0.15}_{\rm stat}{+0.23}_{\rm syst}}$& 62.8 & 22.7 & 14.5 & 0.0 \\ 
                                          &  DOM11& $1.26_{{-0.17}_{\rm stat}{-0.25}_{\rm syst}}^{{+0.18}_{\rm stat}{+0.25}_{\rm syst}}$& 61.6 & 22.1 & 16.3 & 0.0 \\  
                                          &  FR08&  $1.20_{{-0.15}_{\rm stat}{-0.21}_{\rm syst}}^{{+0.15}_{\rm stat}{+0.23}_{\rm syst}}$ & 64.5 & 21.5 & 14.0 & 0.0 \\ 
      \hline
    \end{tabular}
    \tablefoot{The reported $\alpha$ values are obtained using Eq.~\ref{average_alpha} as described in Sect.~\ref{results}. Col. 1: redshift bins. Col. 2: EBL models: FR17 \citep{Franceschini2017, Franceschini2018}, GIL12 \citep{Gilmore2012}, DOM11 \citep{Dominguez2011}, FR08 \citep{Franceschini2008}. Col. 3: EBL normalization, $\alpha$, and related statistical and systematic uncertainties. The systematic uncertainty includes the effect of the injected intrinsic spectral models, the $\sigma$ threshold, and the flux scaling-factor
    (see Sect.~\ref{model_selection}). These effects were added in quadrature. Col. 4, 5, 6, 7: percentage of final PWL, LP, EPWL, and ELP spectral intrinsic model, respectively, averaged over the results from the four different choices of starting model.
    }
\end{table*}
}

\section{Results}
\label{results}

 The EBL scaling factor was obtained for four different EBL models: FR08, DOM11, GIL12, and FR17. Each spectrum considered in the fit includes points retrieved from both the 3FGL and the 3FHL {\it Fermi}-LAT catalogs, whose treatment is explained in Sect.~\ref{data_sample}. The final $\alpha$ values reported in Table~1 correspond to the average among those obtained starting from different initial spectral models:

\begin{equation}
\alpha = \frac{\alpha_{\rm PWL} + \alpha_{\rm LP} + \alpha_{\rm EPWL} + \alpha_{\rm ELP}}{4} {\text  ,}
\label{average_alpha}
\end{equation}

\noindent and the related statistical uncertainties are estimated by:

\begin{equation}
\sigma_{\rm stat} = \sqrt{\frac{\sigma_{\rm PWL}^2 + \sigma_{\rm LP}^2 + \sigma_{\rm EPWL}^2 + \sigma_{\rm ELP}^2}{4}} {\text  ,}
\end{equation}

\noindent where $\sigma_{\rm PWL}$, $\sigma_{\rm LP}$, $\sigma_{\rm EPWL}$, and $\sigma_{\rm ELP}$ are estimated from the likelihood profile, and correspond to a drop of $1/e$ from the maximum.

\begin{figure}[t!]
\includegraphics[width=8.5cm]{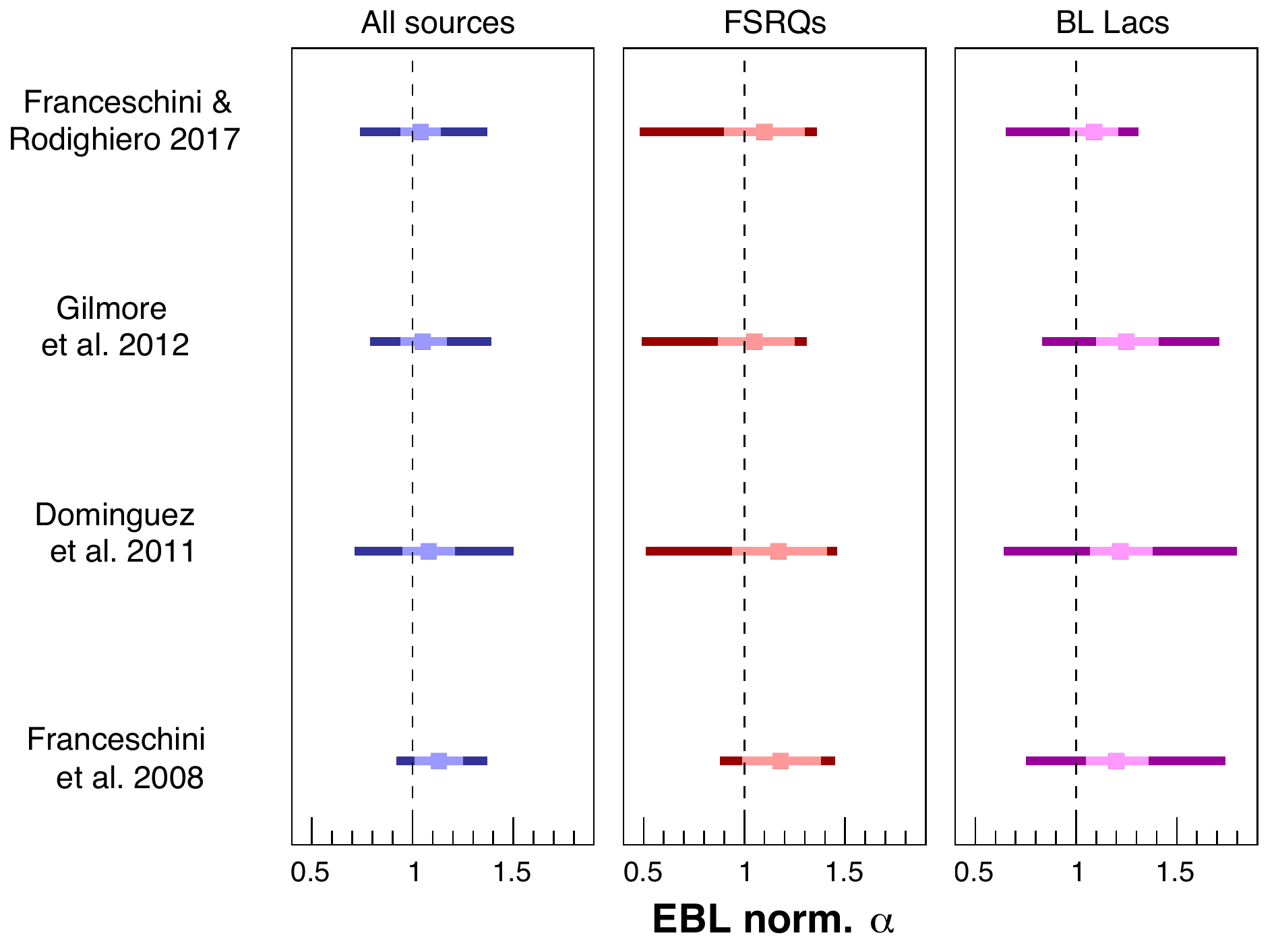}
\caption{The EBL scaling factor obtained analyzing the whole sample (left), FSRQs (center), and BL~Lacs (right)  for FR17, GIL12, DOM11, and FR17 EBL models. Light colors report the statistical uncertainties, while dark colors include also the systematic ones.}
\label{whole_sample_res}
\end{figure}

Two subsamples were analyzed depending on the object class: BL~Lacs and FSRQs. The FSRQ sample contains 157 objects in the redshift range $0.097 \le z \le 2.534$, while the BL~Lac sample contains 299 objects with a redshift $0.032 \le z \le 2.471$.
The other 34 sources include 8 radio galaxies, 24 sources marked as unknown, a narrow-line Seyfert I, and a starburst galaxy.
Figure~\ref{whole_sample_res} reports the results obtained for the whole sample, the BL~Lac sample, and the FSRQ sample, respectively. Results include both statistical and systematic uncertainties; the estimation of the latter is described below.

\begin{figure}[t!]
\includegraphics[scale=0.43]{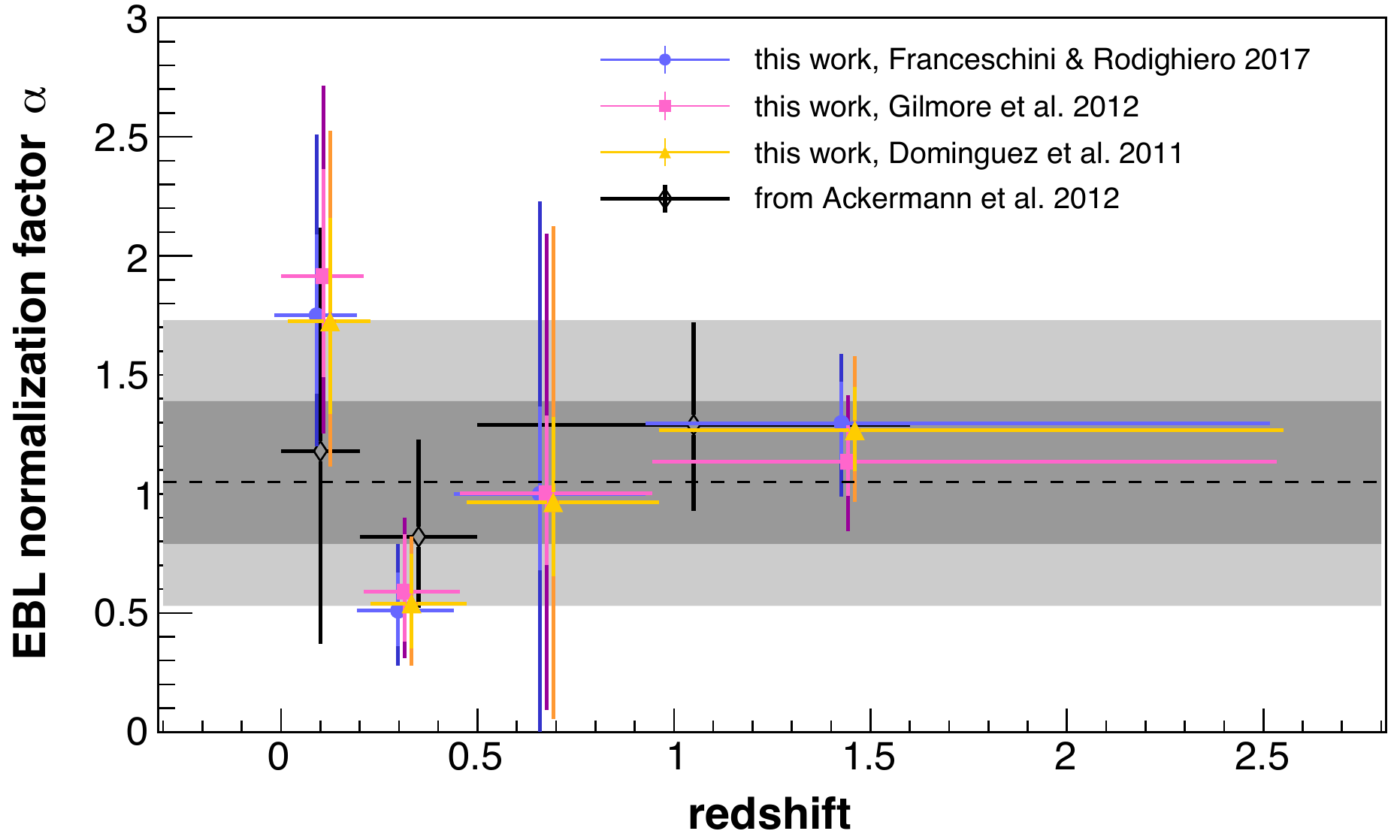}
\caption{The EBL scaling factor obtained for different redshift bins for FR17, GIL12, and DOM11 EBL models. Light colors refer to the statistical uncertainties, while darker colors include also the systematic ones. Comparison with \citep{Ackermann2012} is also shown (gray). The shaded regions show the 1$\sigma$ (dark gray) and 2$\sigma$ (light gray) confidence intervals
obtained for the whole sample, using the GIL12 EBL model and including both statistical and systematic uncertainties.}
\label{binned_res}
\end{figure}

The source sample has also been divided into four redshift bins of similar size: (i) $z \le 0.21$ containing 123 sources; (ii) $0.21 \le z < 0.456$ containing 123 sources; (iii) $0.456 \le z < 0.944$ containing 123 sources, and (iv) $0.944 \le z \le 2.534$ containing 121 sources. Results are reported in Fig.~\ref{binned_res}.

The approach described in this paper is affected by two kinds of systematic errors: (i) the choice of the $\sigma$-threshold (used to switch among the intrinsic spectral models; see Sect.~\ref{model_selection}), and (ii) the injected spectral models which act as a seed for the model-selection procedure. The first is estimated on the whole sample using LP functions as starting $\gamma$-ray spectral models, the GIL12 EBL model, and varying the $\sigma$-threshold from its optimal value ($\sigma=2$) by an amount equal to the standard deviation obtained from the fit of the KS probability (see Sect.~\ref{model_selection}):

\begin{equation}
\begin{split}
\sigma_{\rm syst_{\sigma-thr}} & = \sqrt{\frac{[\alpha_{(2\sigma,{\rm LP})} - \alpha_{(1.8\sigma,{\rm LP})}]^2 + [\alpha_{(2\sigma,{\rm LP})} - \alpha_{(2.2\sigma,{\rm LP})}]^2}{2}}
\\
& = 0.16 {\text  .}
\end{split}
\end{equation}

\noindent The resulting 0.16 is used as a contribution to the systematic uncertainty for all values of $\alpha$.
The choice of injecting a LP function to model the intrinsic spectra does not have a significant impact on the estimation of this uncertainty.

The second source of systematic uncertainty is estimated by fixing the $\sigma$ threshold to its optimal value, and by varying the starting spectral models: 
\begin{equation}
\begin{split}
\sigma_{\rm syst_{mod.sel.}} & =  \left[[ (\alpha_{(\sigma=2.0, {\rm PWL})} - \alpha)^2 + (\alpha_{(\sigma=2.0, {\rm LP})} - \alpha)^2 \right. \\
                               & \left. + (\alpha_{(\sigma=2.0, {\rm EPWL})} - \alpha)^2 + (\alpha_{(\sigma=2.0, {\rm ELP})} - \alpha)^2 ]/4 \right] ^{1/2} {\text  .}
\end{split}
\end{equation}

A source of systematic uncertainty specific to this study is related to the flux scaling factor applied to the 3FGL spectral points (see Sect. \ref{spectral_points}). The two different scaling factors obtained for the steady and the variable samples were applied to all the sources studied in this work. The systematic uncertainty $\sigma_{\rm syst_{flux-corr.}}$ is estimated as the difference between the EBL normalization obtained with the whole-sample scaling factor, and those obtained with the steady- and variable- sample scaling factors.

\noindent Contrary to the $\sigma$ threshold, the uncertainties related to the injected models and flux scaling factor are estimated for each bin, since they may have a different impact depending on the redshift ({\it i.e.}, on the absorption amount due to the EBL photons).
For reference, we obtain \mbox{$\sigma_{\rm syst_{mod.sel.}}=0.01$} and \mbox{$\sigma_{\rm syst_{flux-corr.}}=-0.18+0.28$} with the model of GIL12 applied to the full sample.

Finally, statistic and systematic uncertainties were added in quadrature to obtain the total error:

\begin{equation}
\sigma_{\rm tot} = \sqrt{\sigma_{\rm stat}^2 + \sigma_{\rm syst_{\sigma-thr}}^2 + \sigma_{\rm syst_{mod.sel.}}^2 + \sigma_{\rm syst_{flux-corr.}^2}} {\text  .}
\end{equation}

\section{Discussion}
\label{discussion}

A sample of 490 blazars observed with {\it Fermi}-LAT was used to determine the EBL normalization, $\alpha$, for the EBL models of FR08, DOM11, GIL12, and FR17.
The joint fit of the EBL optical depth and of the intrinsic spectra of the sources was performed allowing any of four possible intrinsic $\gamma$-ray spectral shapes.
A set of intrinsic spectral models is injected at the beginning of the fitting procedure, and the choice for each source is left free to vary according to the best overall fit to the data.
The systematic uncertainties induced by the choice of the initial intrinsic models were estimated for the injected models and the selection criterion. Another source of systematic uncertainty
derives from the flux scaling factor applied to correct the offset between the two catalogs. An important role in the determination of such a factor might be played by the source variability or by the low statistics at higher energies. This systematic uncertainty, specific to this study and not intrinsic to the method itself, is comparable to and often larger than the others.

\begin{figure*}[tp!]
\centering
\includegraphics[scale=0.887]{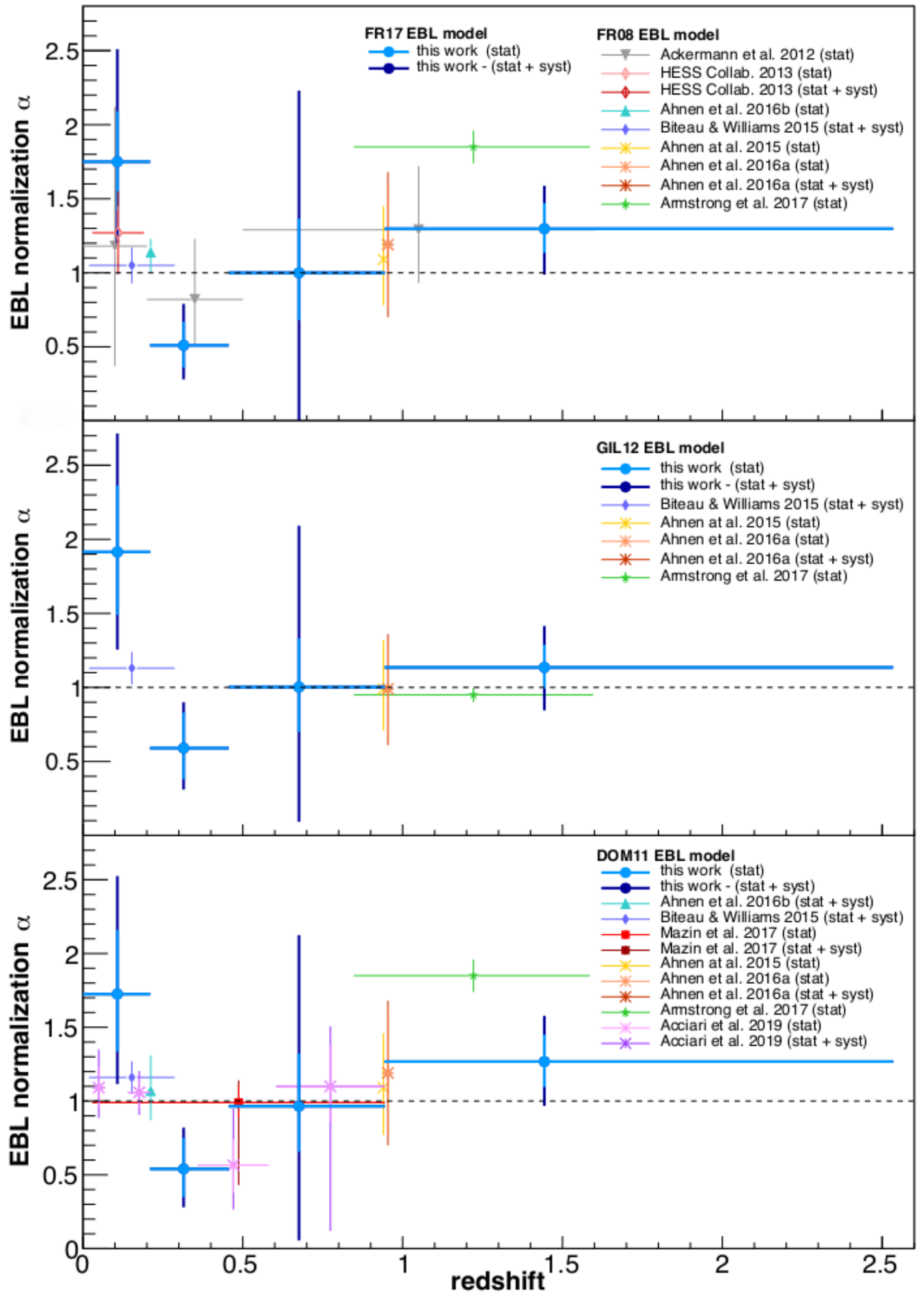}
\caption{Comparison among the results in the literature and those obtained in this work. Comparisons are shown for different EBL models: FR08 and FR17 (top), GIL12 (central), and DOM11 (bottom).}
\label{comparison_results}
\end{figure*}

The analysis performed on the whole sample gives an EBL normalization, $\alpha$, ranging from 1.04 to 1.13, which is compatible with a value of 1 for all the EBL models, with a typical statistical uncertainty of $\sim$10-12\%, and a total uncertainty (statistical + systematic) of $\sim$25-34\%, and up to 39\% for the DOM11 EBL-model.

Sources were divided into BL~Lacs and FSRQs in order to see if internal absorption plays a different role in different object types, possibly biasing the EBL estimation. The values of the scaling factor, $\alpha$, obtained for the two subsamples are perfectly compatible with each other, but the analysis ends up with a significant difference in the final intrinsic $\gamma$-ray spectral models. The intrinsic spectra of BL~Lacs are modeled mainly by PWL, $\sim$80\%, and by only $\sim$16\% of LP and EPWL. The latter percentage rises to $\sim$45\% in the case of FSRQs, which are modeled by a PWL in only $\sim$55\% of sources.
An explanation for this can be found in the fact that FSRQs are characterized by a synchrotron peak located at low energies ($\nu_{\rm s} < 10^{14.5}$\,Hz), while the peak distribution of BL~Lacs is shifted to higher energies by at least one order of magnitude \citep{Giommi2012}. The high-energy component of blazar spectra similarly exhibit a maximum in their SED located at energies $>1$\,TeV and $\le 100\,$GeV for BL~Lacs and FSRQs, respectively. Therefore, the difference in the percentage of intrinsic models of a particular type can be explained by the energy range explored by {\it Fermi}-LAT.

The sample was then split into four redshift bins, each containing a similar number of sources. Within each bin, similar values of the EBL scaling factor were found for the four EBL models.
All results are compatible within $2\sigma$ with the EBL normalization obtained for the whole sample.

Finally, from Table \ref{results_table1}, we note that the percentage of sources preferring PWL and LP changes slightly as the redshift increases. In particular, moving to higher redshift the fraction of PWL models decreases (from $\sim$85 to $\sim$63\%), while the fraction of LP models increases (from $\sim$8 to $\sim$24\%), and similarly for EPWL (from $\sim$6 to $\sim$15\%). The evolution of the fraction of LP models can be understood from the presence of a larger number of FSRQs at higher redshift, with FSRQs peaking at lower energies than BL Lacs.

In Fig.~\ref{comparison_results}, the EBL normalizations obtained in this work are compared with those present in the literature.
In the top panel of Fig.~\ref{comparison_results}, our results obtained with FR17 are compared to existing results obtained with FR08. We note that in FR17, most of the updates in the EBL spectrum with respect to FR08 have modified the CIB peak in the mid- to far-infrared ranges. Since the COB peak provides the EBL photons interacting with $\gamma$-rays detectable by {\it Fermi}-LAT, we expect that changes in FR17 only slightly affect the results obtained using the {\it Fermi}-LAT data, and hence that the results are comparable to those obtained by adopting the FR08 EBL model. The difference between the results obtained with the two models is $\sim$6\% for the first and the second bin, $\sim$10\% for the third bin, $\sim$8\% for the fourth bin, and $\sim$8\% for the whole sample.

The results of this work obtained for the first two bins --- adopting the FR17 EBL model --- are compatible with those obtained for FR08 in \cite{Ackermann2012}, where a smaller sample of 150 blazars was analyzed assuming an intrinsic spectral shape described by an LP function. The results presented here are also consistent with the results of \cite{Hess2013} and of \cite{Mazin2017}. 

Compatible results with this work are seen from the comparison of the EBL scaling factor obtained from the individual blazar PKS 1441+25 \citep{Ahnen2015} and the lensed blazar B2 018+357 \citep{Ahnen2016}.

From Fig. \ref{comparison_results}, one could possibly infer a tension between the results of this work and those (i) in \cite{Ahnen2016b} obtained around $z\sim0.2$, in \cite{Biteau2015} obtained with TeV sources in the redshift range $0.019<z<0.287$, and in \cite{Acciari2019} obtained with GeV-TeV sources in the redshift range $0.14 \le z \le 0.212$; and (ii) in \cite{Armstrong2017}, obtained with $17 \,$ GeV sources in the redshift range $0.847<z<1.596$. In order to investigate if a discrepancy is effectively present, we performed a dedicated analysis in the same redshift ranges as those used in the literature. Results obtained at low redshifts ($0.019 \le z \le 0.287$) are compatible with those shown in \cite{Biteau2015}, differing by $0.2 \sigma$ for the FR08 EBL model, $0.6 \sigma$ for GIL12, and $0.5 \sigma$ for DOM11. Results obtained in the redshift range $0.14 \le z \le 0.212$ for the DOM11 EBL model show a $2.3 \sigma$ tension with respect to those obtained in \cite{Acciari2019}.
At higher redshifts ($0.85 \le z \le 1.596$), the comparison with results in \cite{Armstrong2017} shows a small tension: $1.7\sigma$ for the FR08 EBL model, and $1.6\sigma$ for DOM11. On the contrary, the result of these latter authors for the GIL12 EBL model is in agreement with that presented in this work $(0.8\sigma)$, as well as results presented in \cite{Ahnen2015} and \cite{Ahnen2016}.

Recently, \cite{Abdollahi2018} carried out studies on the EBL using a large sample of 739 AGNs detected with the {\it Fermi}-LAT telescope. These latter authors obtained normalization factors (with $1 \sigma$ confidence range) of $1.30 \pm 0.10$, $1.31 \pm 0.10$, and $1.25 \pm 0.10$ for the GIL12, DOM11, and FR17 EBL model, respectively. These values are compatible with those obtained in this work for the whole sample, considering both statistical and systematic uncertainties.

\section{Conclusion}

This paper exams a methodology to determine the scaling factor for different EBL models, avoiding a priori assumptions on the intrinsic spectra of the sources as much as possible. In fact, the spectral curvature is a critical factor that might affect a reliable estimation of the EBL scaling factor, whose impact may not have been fully quantified thus far since the internal emission and the absorption processes at play in blazars are not fully understood. Not accounting for curvature can lead to an overestimation of the scaling factor, attributing all the absorption to the EBL. On the contrary, an intrinsic curvature that is too accentuated can retrace the EBL absorption, leading to an underestimation of the latter. The approach we use, expanding
on that presented in \cite{Biteau2015}, leaves the intrinsic spectral models free in the fitting procedure, avoiding a fixed intrinsic $\gamma$-ray spectral shape.
This approach is accompanied by a careful study of the systematic uncertainties, possibly the first of its kind, where four different functions are used as candidates to model the intrinsic spectra. 

The EBL scaling factors obtained by analyzing 490 {\it Fermi}-LAT archival spectra with this method are in good agreement with those presented in the literature, and they were obtained with a robust approach.
In this work, the study of the evolution of the EBL appears to be limited by the systematic uncertainties due to model selection. An interesting comparison would consist of applying the approach described in this paper to the larger sample of \cite{Abdollahi2018}, who adopted another method to find the EBL normalization, and to run a dedicated spectral analysis exploiting the full potential of {\it Fermi}-LAT data rather than using archival catalog data.
Moreover, constraints on the EBL evolution can potentially be improved using the {\it Fermi}-LAT Fourth Source Catalog (4FGL),\footnote{\url{https://fermi.gsfc.nasa.gov/ssc/data/access/lat/8yr_catalog/}} or more specifically the 4LAC, which will feature spectral points and redshifts of AGNs. This will allow the unabsorbed part of the spectrum to be better constrained. Finally, the future Cherenkov Telescope Array (CTA) ground-based $\gamma$-ray observatory \citep{ScienceCTA2017}, with its unprecedented sensitivity, will be a crucial instrument for a better understanding of the EBL.

\begin{acknowledgements}
We are grateful for a fruitful exchange with the anonymous referee, which helped to improve the quality of this paper. We thank Paolo Goldoni for feedback on the redshifts of the sources studied in this work. We acknowledge support from the U.S. NASA {\it Fermi}-GI Cycle 9 grant NNX16AR40G. This work was also supported by the {\it Programme National Hautes Energies} with the founding from CNRS (INSU/ INP3/INP), CEA, and CNES. Moreover, we thank U.S. National Science Foundation for support under grants PHY-1307311 and PHY-1707432 as well as P2IO LabEx (ANR-10-LABX-0038) in the framework ``Investissements d'Avenir'' (ANR-11-IDEX-0003-01) managed by the ``Agence Nationale de la Recherche'' (ANR, France).
\end{acknowledgements}

\bibliographystyle{aa}
\bibliography{references}

\end{document}